\documentclass[groupedaddress,showkeys]{revtex4-2}
\usepackage{float}
\usepackage{multirow}
\usepackage{amsmath}
\usepackage{amssymb}
\usepackage{graphicx}
\usepackage{xcolor}
\usepackage{stackrel}
\usepackage{natbib}
\definecolor{tl}{RGB}{0,180,120}
\usepackage[colorlinks,citecolor=tl,linkcolor=red,urlcolor=magenta,unicode=true]{hyperref}

\begin{document}
	\title{Future deceleration due to backreaction in a Universe with multiple inhomogeneous domains}
	
	\author{Ashadul Halder}
	\email{ashadul.halder@gmail.com}
	\affiliation{Department of Astrophysics and Cosmology, S. N. Bose National Centre for Basic Sciences,\\ JD Block, Sector III, Salt Lake, Kolkata-700106, India.}
	
	\author{Shashank Shekhar Pandey}
	\email{shashankshekhar.pandey@bose.res.in}
	\affiliation{Department of Astrophysics and Cosmology, S. N. Bose National Centre for Basic Sciences,\\ JD Block, Sector III, Salt Lake, Kolkata-700106, India.}
	
	\author{A. S. Majumdar}
	\email{archan@bose.res.in}
	\affiliation{Department of Astrophysics and Cosmology, S. N. Bose National Centre for Basic Sciences,\\ JD Block, Sector III, Salt Lake, Kolkata-700106, India.}
	
	\begin{abstract}
		\begin{center}
			\large{\bf Abstract}
		\end{center}

		We formulate a  model of spacetime with inhomogeneous matter distribution in multiple domains. In the context of the backreaction framework using Buchert's averaging procedure, we evaluate the effect of backreaction due to the inhomogeneities on the late time global evolution of the Universe. Examining the future evolution of this universe, we find that it can transit from the presently accelerating phase to undergo future deceleration. The future deceleration is governed by our model parameters. We constrain the model parameters using observational analysis of the Union 2.1 supernova Ia data employing the Markov Chain Monte Carlo method.
		
	\end{abstract}
	\maketitle

\section{Introduction }\label{sec:intro}

The concordance model of cosmology, {\it viz.}, the $\Lambda$CDM model, is based on the Cosmological Principle, which states that the Universe is homogeneous and isotropic and interprets the Universe as an FLRW spacetime.  Cosmological observations like Sloan Digital Sky Survey (SDSS) \cite{Labini_2009} indicate that matter inhomogeneities exist at the scales of super-clusters of galaxies. Studies analyzing large-scale fluctuations in the luminous red galaxy samples \cite{wiegand_scale}  have found substantial (more than three sigmas) divergence from the $\Lambda$CDM mock catalogues on samples as large as $500 h^{-1}$ Mpc. Thus, though the Universe is homogeneous and isotropic at extremely large length scales, the Cosmological Principle doesn't hold at smaller scales, and matter inhomogeneities may have significant consequences up to a length scale of $500 h^{-1}$ Mpc. 

In order to study the effect of inhomogeneities on the Universe's evolution, an averaging procedure is required. The averaging problem was first introduced in general relativity in 1963 \cite{Shirokov1998}, although the proposed method was not covariant. Various categories of averaging techniques
have been since suggested in the literature \cite{Ellis1984, Futamase, Gasperini_2011}. Zalaletdinov introduced a covariant and exact averaging procedure wherein an averaged version of Einstein's equations was attained using a covariant averaging scheme of tensors via bilocal operators \cite{Zalaletdinov1992, Zalaletdinov1993}. Buchert \cite{Buchert, Buchert2001} proposed an averaging procedure in which he simplified the problem restricting it to averaging scalar quantities only. Using Buchert's averaging procedure,
the redshift-distance relation has been shown to be affected by spatial averaging \cite{rasanen1, rasanen2, Koksbang_2019, Koksbang2, Koksbang3, Koksbang4}. Connections between spatial averages and the redshift-distance relation have been further analyzed \cite{Koksbang_2019, Koksbang3}. In this work, we are motivated to use Buchert's averaging procedure as it  provides us a scheme of relating our theoretically calculated quantities (the spatial averages) with observational quantities (the redshift-distance relations) \cite{rasanen1, rasanen2, Koksbang_2019}.

A number of studies have been performed in the context of the Buchert framework to examine the effect of matter inhomogeneities on the large scale cosmological dynamics, and propagation of light and gravitational waves \cite{Coley, Buchert, Buchert2001, Korzynski_2010, Clifton, Skarke, Buchert_2015, Buchert4, Weigand_et_al, Rasanen_2004, wiltshire, Kolb_2006, Koksbang_2019, Koksbang2, Koksbang_PRL, Rasanen_2008, bose, Bose2013, Ali_2017, Pandey_2022, Pandey2023, Koksbang5, Koksbang6, Koksbang7, Koksbang8}.  Although the overall significance of cosmic backreaction on the overall evolution of the real Universe is still debated \cite{Ishibashi_2005}, it is at least possible, in principle, for the backreaction to influence the evolution of the universe \cite{Buchert_2015}. In this work, we employ Buchert's backreaction formalism to examine the fate of the currently accelerating phase of the Universe in the presence of observed matter inhomogeneities at considerably large scales.

Observational evidence establishes the Universe's current acceleration \cite{Perlmutter1998, Riess_1998, Hicken_2009, Marina_Seikel_2009}.
However, the $\Lambda$CDM model is afflicted by certain observational discrepancies, such as the Hubble tension \cite{Riess_2021, Freedman_2021} 
which has attracted a lot of attention recently. The Hubble tension arises from a discrepancy in the inferred value of the Hubble parameter from local measurements compared to that from early Universe physics. It is possible for the backreaction-induced curvature to explain the larger values of the Hubble parameter obtained locally \cite{Heinesen_2020}. The $\Lambda$CDM Universe may end in a future big  freeze, or even in a big rip \cite{Phantom_energy_CALDWELL200223, Phantom_energyPRL, Phantom_energyPRD}  in the presence of phantom dark energy $(\omega_{de} < -1)$. It has been shown that the Buchert formalism in dust universes predicts the dwindling of the present acceleration \cite{Constraints_Räsänen_2006}, and  extrapolation of Buchert's procedure in the context of a two-scale void-wall
toy model  of the Universe leads to the possibility of avoiding the future big  freeze or a possible future big rip in a phantom dark energy model \cite{bose, Bose2013, Ali_2017}. 

The motivation for the present study is to re-examine the late-time
evolution of the presently accelerating Universe by employing the
Buchert backreaction formalism in the context of a more realistic model mimicking our actual Universe with observed inhomogeneities extending up to considerably large scales.  In this work, we consider a multiple subregion model of the spacetime with each subregion having a distinct set of parameters characterizing its evolution. We apply the Buchert averaging procedure
against the backdrop of such a model in order to investigate the global
evolution during the present era extended to the future. It is important to analyse the future evolution ensuing from such a study,  as a transition from present acceleration to deceleration in the near future may have immediate consequences for our present Universe. We confront this scenario against observational results by performing Markov Chain Monte Carlo analysis using the Union 2.1 supernova Ia data \cite{union} to determine our model parameters' best fit and optimum values.

The paper is organized as follows. We briefly introduce Buchert's backreaction formalism in (\autoref{sec:buchert}). In (\autoref{sec:our model}), we introduce our model of inhomogeneities in multiple subregions. In (\autoref{sec:results}), we present our multi-domain model's theoretical analysis leading to the Universe's predicted late-time evolution.  In (\autoref{sec:obs}) we use the Union 2.1 supernova Ia data to constrain our model parameters. We summarise our main results in (\autoref{sec:conclusions}).

\section{Buchert's backreaction formalism}\label{sec:buchert}
In the Buchert formalism, Einstein equations are decomposed into dynamical equations for scalar quantities. Proper volume averages are defined by the averages on flow-orthogonal spatial hypersurfaces (vorticity is assumed zero). This procedure leads to the Buchert equations with a kinematical backreaction term \cite{Buchert, Buchert2001, buchert_rasanen}. In Buchert's averaging scheme for scalars, averages of scalar quantities on flow-orthogonal spatial hypersurfaces are defined as \cite{Weigand_et_al}
\begin{equation} \label{eq:favg}
\langle f(t, x^i)\rangle_D := \frac{\int_D d^{3}x \sqrt{det(g_{ij})} f(t, x^i) }{\int_D d^{3}x \sqrt{det(g_{ij})} } \, , 
\end{equation}
where D is a spatial domain. The volume of such a domain is given by,
\begin{equation}\label{eq:VD}
    V_D(t) := \int_D d^3x  \sqrt{det(g_{ij})} \, .
\end{equation}
The normalized dimensionless effective volume scale factor $a_D$ is defined by
\begin{equation}\label{eq:aD}
    a_D (t) := \left(\frac{V_D(t)}{V_{D_0}}\right)^{1/3} \, , 
\end{equation}
which is normalized by the volume $V_{D_0}$ of the domain $D$ at some reference time $t_0$, which we can take as the present time.

Spatially averaging the Raychaudhuri equation, the Hamiltonian constraint and the continuity equation, one obtains Buchert's equations, which are, respectively,
\begin{equation}\label{eq:Rayeqn}
    3\frac{\ddot{a}_D}{a_D} = -4\pi G\langle\rho\rangle_D + Q_D ,
\end{equation}
\begin{equation}\label{eq:HamiltonianCeqn}
    3H_D^2 = 8\pi G\langle\rho\rangle_D - \frac{1}{2}\langle R\rangle_D - \frac{1}{2}Q_D  ,
\end{equation}
\begin{equation}\label{eq:continuity}
    \partial_t\langle\rho\rangle_D + 3H_D\langle\rho\rangle_D = 0 \, , 
\end{equation}
where local averaged matter density $\langle\rho\rangle_D $,
averaged spatial Ricci scalar $\langle R\rangle_D$ and the Hubble parameter $H_D$ are domain dependent and are functions of time. $Q_D$ is called the kinematical backreaction term which evaluates the averaged effect of the inhomogeneities in the domain $D$ and is defined as
\begin{equation}\label{eq:QD}
    Q_D := \frac{2}{3}\left(\langle \theta^2\rangle_D-\langle\theta\rangle_D^2\right) - 2\langle\sigma^2\rangle_D \, ,
\end{equation}
where $\theta$ is the local expansion rate and $\sigma^2 := \frac{1}{2}\sigma^i_j\sigma^j_i$ is the shear-scalar. $Q_D$ is zero for a FLRW-like domain. $Q_D$ and $\langle R\rangle_D$ are inter-related by the equation:
\begin{equation}\label{eq:condition}
    \frac{1}{a_D^2}\partial_t(a_D^2\langle R\rangle_D) + \frac{1}{a_D^6}\partial_t(a_D^6Q_D) = 0 \, .
\end{equation}
(\autoref{eq:condition}) couples the time evolution of the kinematical backreaction term with the time evolution of averaged intrinsic curvature. This coupling, denoting deviation of the spatial curvature term from being proportional to $1/a^2$, along with the kinematical backreaction term $Q_D$ signifies the departure from FLRW-cosmology. 

We now adopt a specific approach within the Buchert formalism, in which 
 ensembles of disjoint regions are considered to represent the global domain
 \cite{Buchert_2015, Buchert4, Weigand_et_al, Rasanen_2004, wiltshire, Kolb_2006, Koksbang_2019, Koksbang2, Koksbang_PRL, Rasanen_2008, bose, Bose2013, Ali_2017, Pandey_2022, Pandey2023, Koksbang5, Koksbang6, Koksbang7, Koksbang8}.  
 Here, the domain $D$ is partitioned into non-interacting subregions $\mathcal{F}_l$ composed of elementary space entities $\mathcal{F}_l^{(\alpha)}$. Mathematically, we can represent the global domain $D$ as $D = \cup_l\mathcal{F}_l$, where each subregion can be represented as $\mathcal{F}_l = \cup_{\alpha}\mathcal{F}_l^{(\alpha)}$ and the elementary space entities $\mathcal{F}_l^{(\alpha)}$ are all distinct, $\mathcal{F}_l^{(\alpha)}\cap\mathcal{F}_m^{(\beta)} = \emptyset$ for all $\alpha\neq\beta$ and $l\neq m$. The average of any scalar function $f$ on the domain $D$ is given by,
\begin{equation}\label{eq:fD_sum}
\begin{split}
    \langle f \rangle_D &= V_D^{-1}\int_D f  \sqrt{det(g_{ij})} d^{3}x \\
    &= \sum_l V_D^{-1}\sum_{\alpha}\int_{\mathcal{F}_l^{(\alpha)}} f  \sqrt{det(g_{ij})} d^{3}x \\
    &= \sum_l \frac{V_{\mathcal{F}_l}}{V_D}\langle f\rangle_{\mathcal{F}_l} = \sum_l \lambda_l\langle f\rangle_{\mathcal{F}_l} \, , 
\end{split}   
\end{equation}
where $\lambda_l = V_{\mathcal{F}_l}/V_D$ is the volume fraction of the subregion $\mathcal{F}_l$ such that $\sum_l \lambda_l = 1$ and $\langle f\rangle_{\mathcal{F}_l}$ is the average of $f$ on the subregion $\mathcal{F}_l$. The above equation governs the averages of scalar quantities $\rho$, $ R$ and $H_D$. But due to the presence of the $\langle\theta\rangle_D^2$ term, $ Q_D$ does not follow the above equation. Instead, the equation for $Q_D$ is
\begin{equation}\label{eq:QD_sum}
    Q_D = \sum_l \lambda_lQ_l + 3\sum_{l\neq m}\lambda_l\lambda_m(H_l-H_m)^2 \, , 
\end{equation}
where $Q_l$ and $H_l$ are quantities having the same form in the subregion $\mathcal{F}_l$ as $Q_D$ and $H_D$ have in the domain $D$ \cite{Weigand_et_al}.

We can also define scale factor $a_l$ for the individual subregions in the same way as $a_D$ has been prescribed for the domain $D$. Since, the domain $D$ comprises the different subregions $\mathcal{F}_l$ and all these subregions are disjoint, therefore $V_D = \sum_l V_{\mathcal{F}_l}$, which results in $ a_D^3 = \sum_l a_l^3$. Twice  differentiating this relation with respect to foliation time gives us,
\begin{equation}\label{eq:aD_sum}
    \frac{\ddot{a}_D}{a_D} = \sum_l\lambda_l\frac{\ddot{a}_l(t)}{a_l(t)}+\sum_{l\neq m}\lambda_l\lambda_m(H_l -H_m)^2 \, .
\end{equation}

\section{Our multiple subregions model}\label{sec:our model}

Several recent analyses have been carried out within the Buchert formalism with one under-dense and one over-dense subdomain \cite{Koksbang2, Koksbang_PRL,  bose, Bose2013, Ali_2017, Pandey_2022}. The assumption of just
two different density domains represent an oversimplification in the context of the real Universe, as the actual density profile varies across a spectrum from regions of
very low density to those of high density. Hence, to construct a more realistic cosmological model, one needs to consider a larger number of sub-domains with distinct evolution profiles. Since some earlier studies based
on two-domain models have predicted a future deceleration of the 
universe \cite{bose, Bose2013, Ali_2017},
it is interesting to study the future evolution of the universe in the context of a more realistic model having a large number of subdomains with distinct
parameters.

In the context of the backreaction framework employed in this work, we consider a model of the Universe in which the domain $D$ of interest comprises multiple subregions. The multiple subregions in our model can be categorized into two types of regions - (i) overdense regions which are closed dust-only FLRW regions with positive curvature and a deceleration parameter $q_o = -\ddot{a_o}/a_o H_o^2 > 0.5$, and (ii) underdense regions which are flat (zero intrinsic curvature) FLRW regions and having smaller density (as compared to the overdense regions). So, there are in total $n$ subregions in our model, $i$ number of them are overdense and $(n-i)$ of them are underdense.

The scale factor and time of the $i^{\rm th}$ overdense region evolve with development angle $\phi$ of the overdense region as \cite{Ali_2017, weinberg},
\begin{eqnarray}
	a_{o_i} &=& \dfrac{q_{o_i}}{2q_{o_i}-1} (1-\cos{\phi}),\label{eq:ao}\\
	t_i &=& \dfrac{q_{o_i}}{2q_{o_i}-1} \left(\phi-\sin{\phi}\right),\label{eq:t}
\end{eqnarray}
where $q_{o_i}$ is the deceleration parameter of the $i^{\rm th}$ overdense region. The scale factor of the $i^{\rm th}$ under-dense region is taken to evolve as a function of time t given by
\begin{equation}
	a_{u_i} = c_{u_i} t^{\beta_i}
	\label{eq:au}
\end{equation}
In the above expression, $c_{u_i}$ and $\beta_i$ are constant, which determines the time evolution of the $i^{\rm th}$ under-dense subregion. $\beta_i$ varies from 2/3 to 1 to denote any behaviour ranging from a  matter-dominated region ($ \beta_i = 2/3$) up to an accelerating region ($ \beta_i  > 1 $).

Now, applying (\autoref{eq:ao} - \autoref{eq:au}) in \autoref{eq:aD_sum}, the expression of the global acceleration takes the form	
	\begin{eqnarray}
    \frac{\ddot{a}_{\mathcal{D}}}{a_{\mathcal{D}}}&=&\left(\sum_i{-\lambda_{o_i} q_{o_i} H_{o_i}^2}\right)+\left(
\sum_j{\lambda_{u_j}\dfrac{\beta(\beta-1)}{t^2}}\right) \nonumber\\
&&+\left(\sum_{k}\sum_{l}{ \lambda_k \lambda_l \left(H_l-H_k\right)^2}\right) \label{eq:acc}.
\end{eqnarray}
Here $\lambda_{o_i}$ is the volume fraction of the $i^{th}$ overdense region, $\lambda_{u_j}$ is the volume fraction of the $j^{th}$ overdense region, $H_{o_i}$ is the Hubble parameter of the $i^{th}$ overdense region, $\lambda$ is the set of all $\lambda_{o_i}$ and $\lambda_{u_i}$ and $H$ is the set of all $H_{o_i}$ and $H_{u_i}$. The total volume fraction of all the under-dense regions, i.e. $\sum_i{\lambda_{u_i}}$ is given by $\lambda_{u}$. Similarly, $\lambda_{o}$ is the total volume fraction of all the over-dense regions. Clearly, $\lambda_{o} + \lambda_{u} = 1$.
	
The volume fraction of the $i^{th}$ under-dense subregion can be written as,
\begin{eqnarray}\label{eq:lmbda_relation}
    \lambda_{u_i} &=& \frac{V_{u_i}}{V_\mathcal{D}}= \frac{a^3_{u_i}V_{{u_i},0}}{a^3_\mathcal{D}V_{\mathcal{D},0}} = \lambda_{{u_i},0}\frac{a^3_{u_i}}{a^3_\mathcal{D}}\nonumber\\
    &=& \lambda_{{u_i},0} \frac{c^3_{u_i} t^{3\beta_i}}{a^3_\mathcal{D}} = \lambda_{{u_i},0} \left(\dfrac{t}{t_0}\right)^{3\beta_i}\left(\dfrac{a_{\mathcal{D},0}}{a_{\mathcal{D}}}\right)^3,
\end{eqnarray} 
where $t_0$ is a reference time generally taken to be present time, $V_{{u_i},0}$ is the volume of the $i^{\rm th}$ under-dense subregion at the present time, $\lambda_{{u_i},0}$ is the volume fraction of the $i^{\rm th}$ under-dense subregion at the present time, $V_{\mathcal{D},0}$ is the present time value of the volume and $a_{\mathcal{D},0}$ is the present time value of the scale factor of the domain $D$ of interest. The present time value of $\lambda_{o}$ and $\lambda_{u}$ are given by $\lambda_{{o},0}$ and $\lambda_{{u},0}$ respectively, which may be
taken to be $0.09$ and $0.91$ \cite{Weigand_et_al}. In the present analysis, we assume the distribution of $\lambda_{{u_i},0}$ across the $i$ underdense subregions to follow a Gaussian profile within the allowed range of $\beta_i$, given by
\begin{equation}
    \lambda_{{u_i},0} = \dfrac{N_u}{\sigma_u \sqrt{2 \pi}} e^{-(\beta_i-\mu_u)^2/2\sigma_u^2},\label{eq:gauss_u}
\end{equation}
where $N_u$ is a normalization constant, such that, $\lambda_{{u},0}$ which is the total volume fraction at the present time of all the under-dense regions is 0.91 (i.e. $\sum_i{\lambda_{{u_i},0}}=0.91$), $\mu_u$ is the mean value of $\beta_i$ and $\sigma_u$ is the standard deviation of $\beta_i$. The present time volume fractions of the over-dense regions are considered to follow Gaussian distributions within the allowed range of $q_o$, given by
\begin{equation}
    \lambda_{{o_i},0} = \dfrac{N_o}{\sigma_o \sqrt{2 \pi}} e^{-(q_{o_i}-\mu_o)^2/2\sigma_o^2}\label{eq:gauss_o}.
\end{equation}
where $N_o$ is a normalization constant, such that, $\lambda_{{o},0}$ which is the total volume fraction at the present time of all the over-dense regions summed over, $\sum_i{\lambda_{{o_i},0}}=0.09$, $\mu_o$ is the mean value of $q_{o_i}$ and $\sigma_o$ is the standard deviation of $q_{o_i}$. As mentioned earlier, $q_{o_i}$ ranges  from $0.51$ to $1$. $\lambda_{o_i}$ which is the volume fraction of the $i^{th}$ over-dense subregion at a time $t$ is related to $\lambda_{{o_i},0}$ by the relation 
\begin{equation}\label{eq:lambda_o_i}
    \lambda_{o_i} = \lambda_{{o_i},0}\left(\dfrac{1-\sum_i \lambda_{u_i}}{1-\sum_i \lambda_{{u_i},0}}\right),
\end{equation}

\section{Late time evolution of the Universe}\label{sec:results}

In our analysis, there are a total of $n$ subregions. In (\autoref{eq:gauss_o}), for a given value of $\mu_o$ and $\sigma_o$, $q_{o_i}$ can take $i$ number of values in the allowed range, where $i$ is the index number of the overdense subregions. For each value of $q_{o_i}$, there is a corresponding value of $\lambda_{{o_i},0}$, such that $\sum_i{\lambda_{{o_i},0}}=0.09$, and (\autoref{eq:lambda_o_i}) gives us $\lambda_{o_i}$ for each $i^{th}$ overdense subregion.  (\autoref{eq:ao}) and (\autoref{eq:t}) give us the scale factor $a_{o_i}$ and time $t$ for  each $i^{th}$ overdense subregion, corresponding
to the value of $q_{o_i}$, and from these, $H_{o_i}$ can be calculated. Similarly, in the case of underdense subregions, in (\autoref{eq:gauss_u}), for a given value of $\mu_u$ and $\sigma_u$, $\beta_i$ can take $i$ 
number of values in the allowed range, where $i$ is the no. of underdense subregions. For each value of $\beta_i$, there is a corresponding value of $\lambda_{{u_i},0}$, such that $\sum_i{\lambda_{{u_i},0}}=0.91$. Once we have $\lambda_{{u_i},0}$, we can use (\autoref{eq:lmbda_relation}) to get $\lambda_{u_i}$. For each $i^{th}$ underdense subregion, (\autoref{eq:au}) gives us the value of scale factor $a_{u_i}$ for the corresponding value of $\beta_i$, and from this, $H_{u_i}$ can be calculated. We can then use (\autoref{eq:aD_sum}) to calculate the scale factor $a_D$ (as a function of time) of the domain of interest $D$ and from it the Hubble parameter $H_D$ of $D$. 

For our multiple subregions model, (\autoref{eq:QD_sum}) effectively becomes,
\begin{equation}\label{eq:QDsum2}
    Q_{D} = \sum_i Q_{o_i} + \sum_j Q_{u_j} + 3\sum_{l\neq m}\lambda_l\lambda_m(H_l-H_m)^2   \, , 
\end{equation}
where $\lambda$ is the set of all $\lambda_{o_i}$ and $\lambda_{u_i}$ and $H$ is the set of all $H_{o_i}$ and $H_{u_i}$. Since (\autoref{eq:condition}) couples the time evolution of the backreaction term $Q_D$ with the time evolution of the averaged 3-Ricci scalar curvature, it is also applicable for our subregions. Thus, the time evolution of $Q_{o_i}$ and $Q_{u_j}$ is coupled to the evolution of the averaged 3-Ricci scalar curvature of the respective subregions. However, one can choose the curvatures of the individual sub-regions in such a way that the $Q_{o_i}$ and $Q_{u_j}$ terms for these sub-regions become effectively zero \cite{Weigand_et_al, Wiltshire_2007}. This is done by taking the curvature of our underdense region to be zero, i.e., our underdense region is flat. On the other hand, we have assumed our overdense region to have Friedmann-like $a_u^{-2}$ constant curvature term. These assumptions along with (\autoref{eq:condition}) results in, $Q_{o_i} = 0$ and $Q_{u_j} = 0$. The stipulation to FLRW is an approximate assumption governing our present model (in the more general case, the sub-domains may not necessarily be FLRW regions). As seen from (\autoref{eq:QDsum2}), the global backreaction is the sum of three terms. In our approach we have assumed the underdense region to have zero curvature, and the overdense region to have constant curvature, thereby making the first two terms in (\autoref{eq:QDsum2}) to vanish. Hence, in this case, the global backreaction is governed by only the interplay of the sub-domain Hubble evolutions
and volume fractions (third term of \autoref{eq:QDsum2}). On the other hand, if the sub-domains are endowed with dynamical curvature, there could be other intricate effects arising through backreaction.

\begin{figure*}
    \centering
    \begin{tabular}{cc}
	\includegraphics[width=0.42\textwidth]{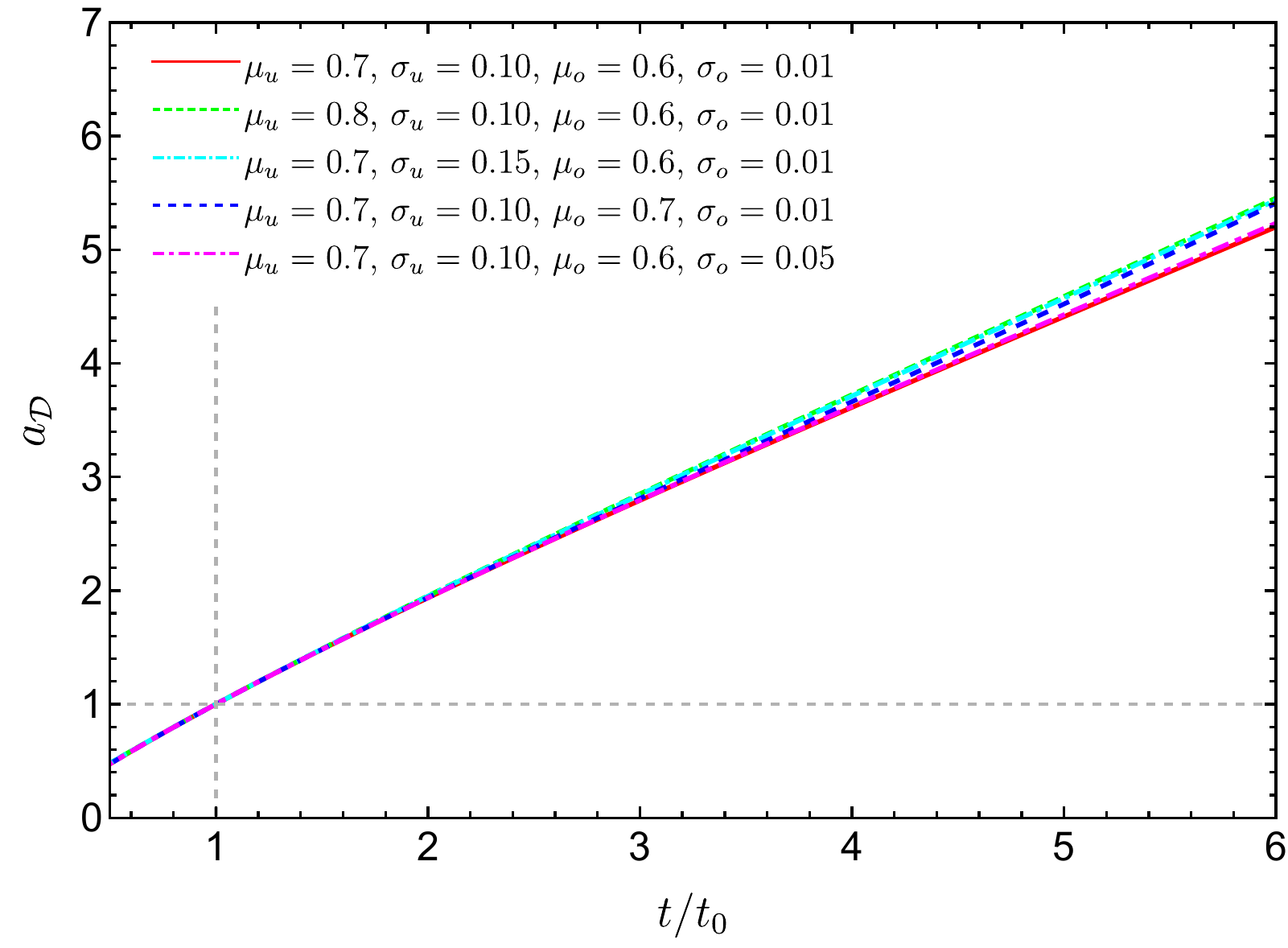}&
	\includegraphics[width=0.45\textwidth]{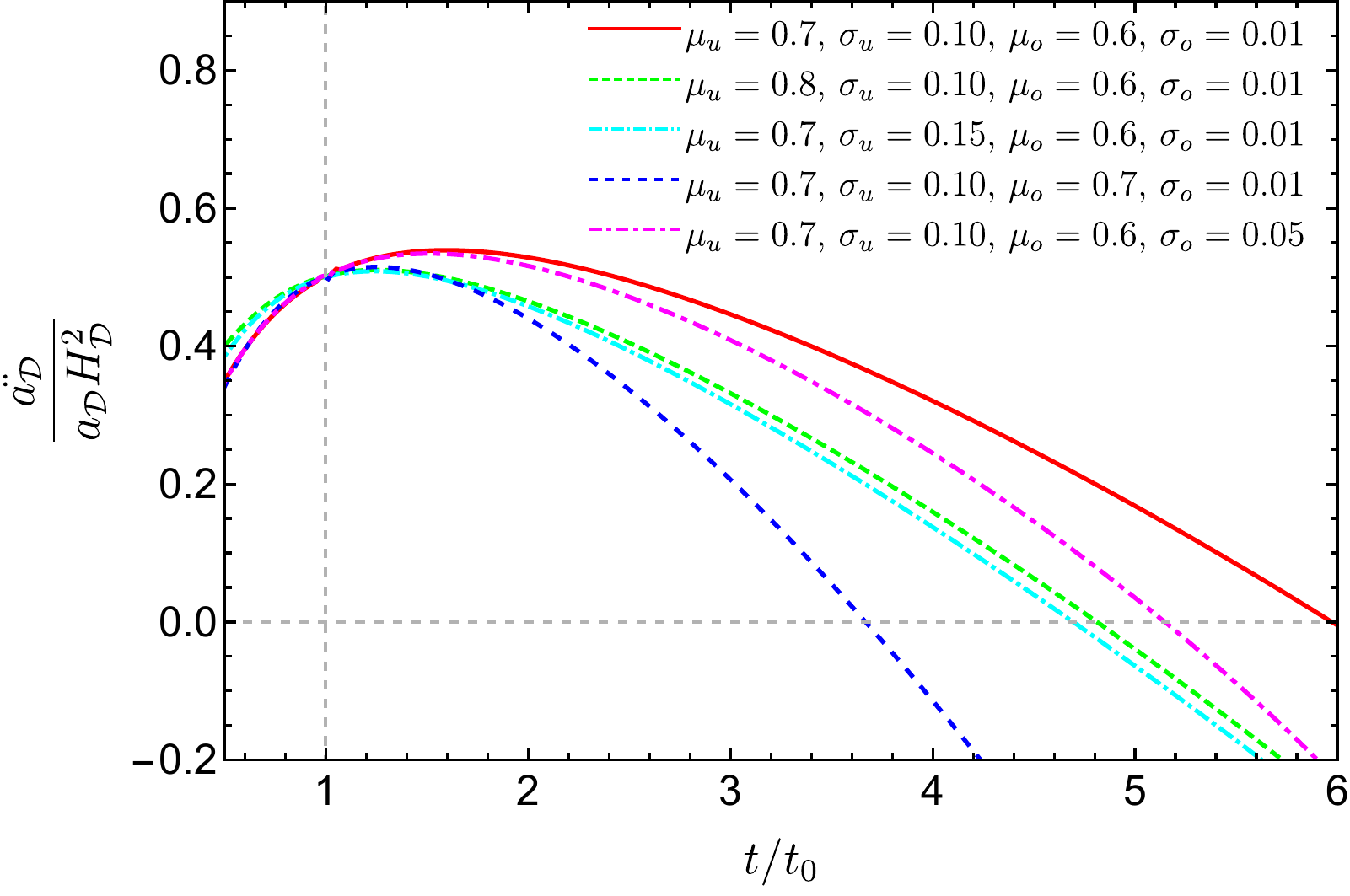}\\
	(a)&(b)\\
    \end{tabular}
    \caption{\label{fig:aD_qD} Evolution of (a) the global scale factor $a_{\mathcal{D}}$, (b) the global acceleration parameter $\frac{\ddot{a_{\mathcal{D}}}}{a_{\mathcal{D}} H_{\mathcal{D}}^2}$, for different sets of model parameters namely, $\mu_o$, $\sigma_o$, $\mu_u$ and $\sigma_o$.}
\end{figure*}
	
In our calculations, we consider one hundred under-dense and one hundred over-dense sub-domains.  These sub-domains are characterized by the respective volume fractions, $\lambda_{o_i}$ and $\lambda_{u_i}$, distributed using a Gaussian profile among these sub-domains (\autoref{eq:gauss_u} and \autoref{eq:gauss_o}). $\mu_o$ and $\sigma_o$ are the mean and standard deviation respectively for the Gaussian profile of overdense regions. In a Gaussian distribution, the mean is also the most frequent observation. Therefore, this means that out of the 100 overdense sub-domains, $\mu_o$ is the most frequent value of $q_{o_i}$. $\sigma_o$ is the distribution's standard deviation, which governs the distribution's width about the mean value. Similarly, $\mu_u$ and $\sigma_u$ are the mean and standard deviation of the Gaussian profile for underdense regions. Our underdense regions are characterized by parameters $\beta_i$ and these vary within the range $2/3 < \beta_i < 1$. This range for $\beta_i$ has been taken to ensure a wide range of underdense subregions is present in our model to  mimic a variety of underdense regions that may be present in the Universe. $\mu_u$ is the most frequent value of $\beta_i$ among the 100 underdense regions. Here, $\sigma_u$ governs the width of the distribution about a given $\mu_u$.


\par In \autoref{fig:aD_qD}(a) and \autoref{fig:aD_qD}(b), the scale factor $a_{\mathcal{D}}$ and the acceleration of the Universe $\frac{\ddot{a_{\mathcal{D}}}}{a_{\mathcal{D}} H_{\mathcal{D}}^2}$ are plotted respectively, for different sets of model parameters \emph{viz.}, $\mu_o$, $\sigma_o$, $\mu_u$ and $\sigma_o$. \autoref{fig:aD_qD}(a) is plotted by fixing the present value (i.e., value at $t = t_0$) of the scale factor $a_{\mathcal{D}}$ to be 1. \autoref{fig:aD_qD}(b) is plotted using the present value (i.e., value at $t = t_0$) of global acceleration parameter $\frac{\ddot{a_{\mathcal{D}}}}{a_{\mathcal{D}} H_{\mathcal{D}}^2}$ to be 0.55 which is obtained using the values of cosmological parameters from Planck 2018 results \cite{Planck}. From \autoref{fig:aD_qD}(b) it can be seen that the acceleration parameter begins to fall off beyond the present era ($t = t_0$), and the Universe transits to a decelerating phase $\left(\frac{\ddot{a_{\mathcal{D}}}}{a_{\mathcal{D}} H_{\mathcal{D}}^2} < 0\right)$ at a subsequent time. This result is in agreement  with observations made in Ref.~\cite{Constraints_Räsänen_2006}, where it was shown that accelerated expansion from backreaction cannot go on forever in the context of dust Universe.
It may be noted that the main focus of the work \cite{Constraints_Räsänen_2006} is to understand whether backreaction could provide a viable means of the current acceleration of the Universe. On the other hand, the aim of our present work is to understand how the Universe evolves in the future, considering it is currently accelerating.

From our results, it can be seen that for higher values of the model parameters, namely, $\mu_o$, $\sigma_o$, $\mu_u$ or $\sigma_o$, the acceleration parameter $\frac{\ddot{a_{\mathcal{D}}}}{a_{\mathcal{D}} H_{\mathcal{D}}^2}$ falls more rapidly with time in comparison to lower values of parameters. Such behaviour is observed since the most frequent values of $\beta_i$ and $q_{o_i}$ in the respective distribution increases  for higher values of $\mu$ ($\mu_u$ and $\mu_o$). Therefore, there are now more underdense and overdense subdomains in the distribution with these higher values of $\beta_i$ and $q_{o_i}$ respectively. Further, increasing the values of $\sigma$ ($\sigma_u$ and $\sigma_o$) results in the distribution of underdense and overdense sub-regions becoming wider around the mean values $\mu_u$ and $\mu_o$ respectively (see \autoref{eq:gauss_u} and \autoref{eq:gauss_o}). As a consequence, the effect of subregions having higher values of $\beta$ (for underdense regions) and $q_o$ (for overdense regions) become prominent in the global dynamic of the Universe, which leads to  rapid fall of the acceleration parameter $\frac{\ddot{a_{\mathcal{D}}}}{a_{\mathcal{D}} H_{\mathcal{D}}^2}$. Note further that higher values of $\beta$ produce a larger acceleration in the early evolution (compare the green dashed line and the red solid line of \autoref{fig:aD_qD}). The green dashed line has a larger value of $\mu_u$ and hence, the most frequent value of $\beta_i$ in the distribution is larger. However, the future acceleration falls off more rapidly due to such higher values of $\beta$.

\begin{figure}
	\centering{}
    \includegraphics[width=0.6\textwidth]{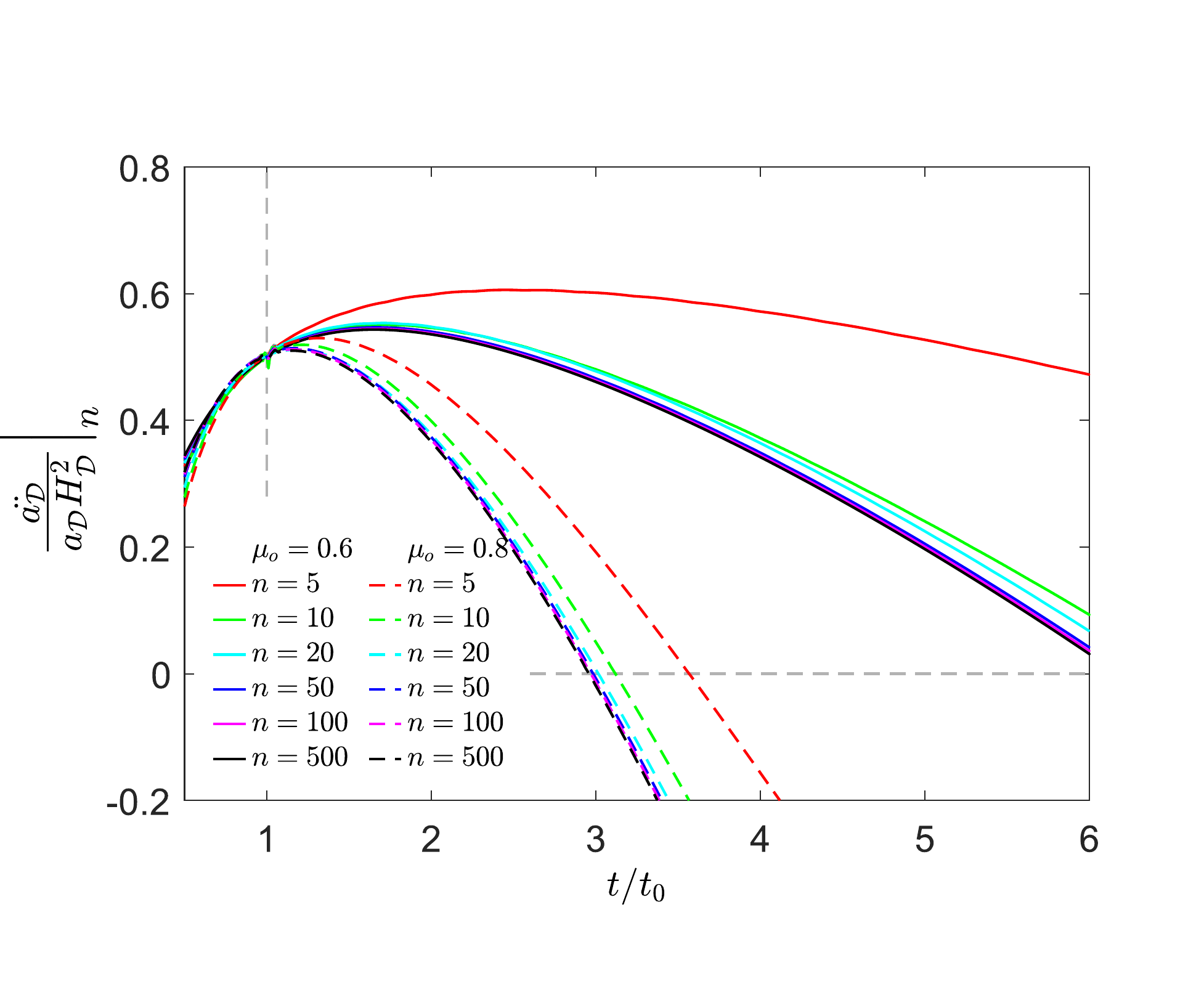}
    \caption{\label{fig:qD_n} Variation of the evolution of global acceleration $\frac{\ddot{a_{\mathcal{D}}}}{a_{\mathcal{D}} H_{\mathcal{D}}^2}$ for different  values of $n$ with $\mu_o=0.6$ (solid lines) and $\mu_o=0.8$ (dashed lines). The other parameters are kept fixed at $\mu_u=0.67$, $\sigma_u=0.1$ and $\sigma_o=0.01$}
\end{figure}

In \autoref{fig:qD_n}, the evolution of the acceleration parameter $\left(i.e.\,\frac{\ddot{a_{\mathcal{D}}}}{a_{\mathcal{D}} H_{\mathcal{D}}^2}\right)$ is plotted for different  values of $n$ with $\mu_o=0.6$ (solid lines) and $\mu_o=0.8$ (dashed lines). Here other parameters are kept fixed at $\mu_u=0.67$, $\sigma_u=0.1$ and $\sigma_o=0.01$. From this figure, it can be observed that the future acceleration falls off more rapidly with the increasing
number of sub-regions. However, if one raises the number of sub-domains 
further,  the evolution of the acceleration parameter tends toward a limiting profile depicted here for the case of $n=500$. 
	
\begin{figure*}
    \centering
    \begin{tabular}{cc}
        \includegraphics[width=0.5\textwidth]{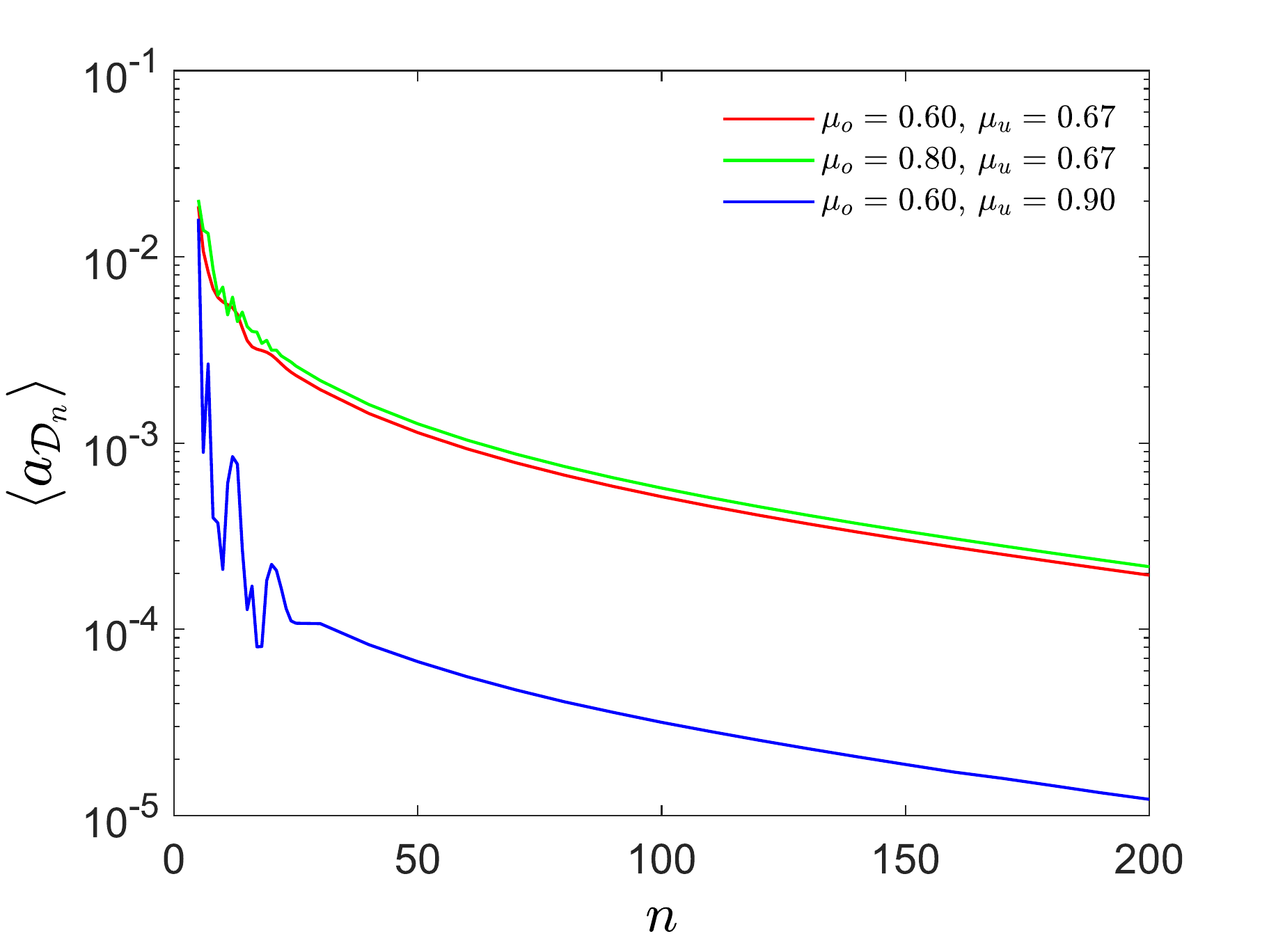}&
	\includegraphics[width=0.5\textwidth]{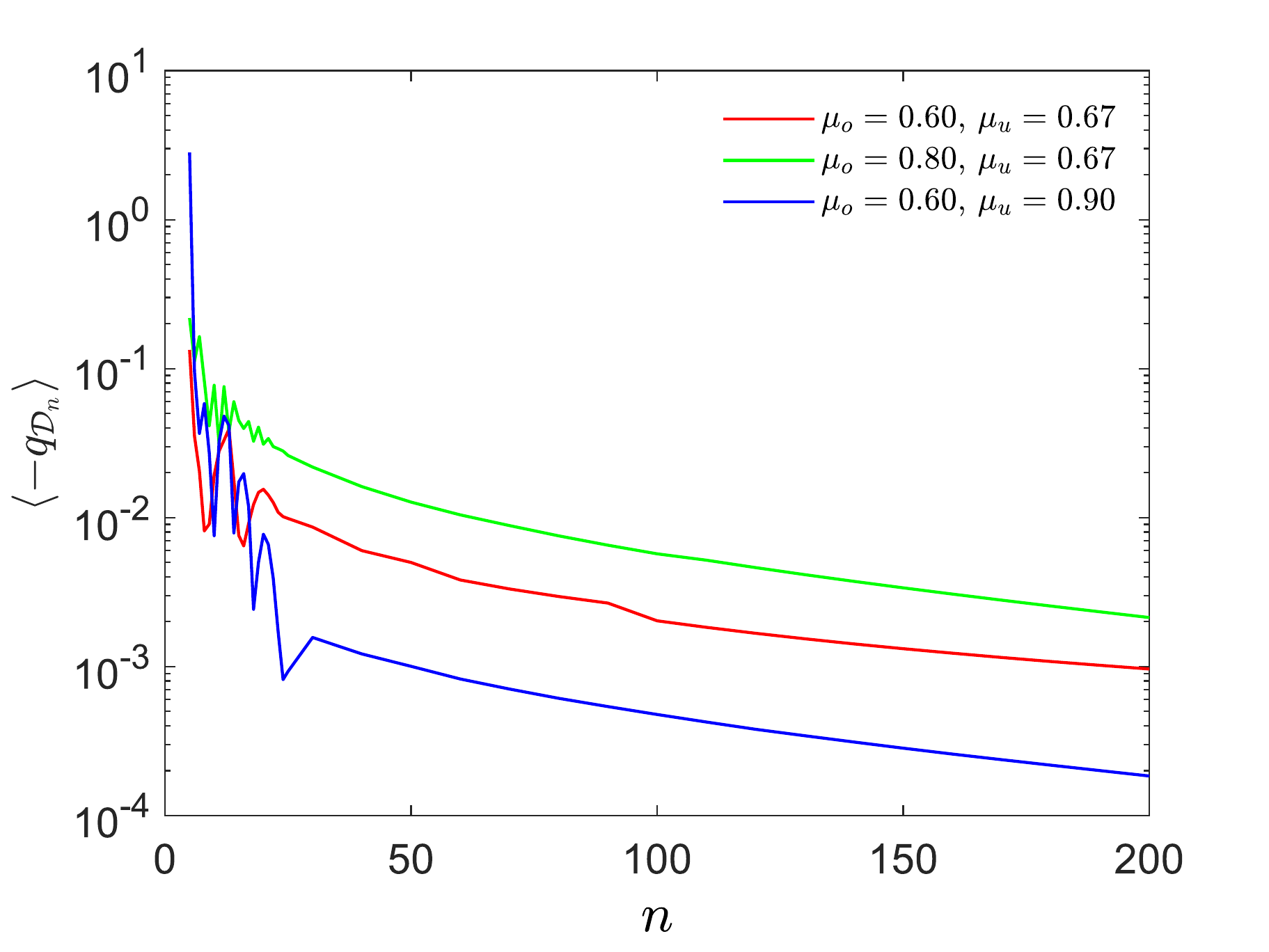}\\
	(a)&(b)\\
    \end{tabular}
    \caption{\label{fig:aD_qD_n_n} Variation of the time averaged (a)  scale factor $\left<a_{\mathcal{D}_n} \right>$ and (b) acceleration parameter $\left<-q_{\mathcal{D}_n} \right>$ with $n$ for different values of $\mu_u$ and $\mu_o$, where the other two parameters are kept fixed at $\sigma_u=0.1$ and $\sigma_o=0.01$.}
\end{figure*}
	
For a deeper investigation into the variation of the global evolution versus the number of sub-domains, we introduce two parameters defined as,
\begin{eqnarray}
    \left<a_{\mathcal{D}_n} \right> &=&  \dfrac{1}{\sum{i}} \sum_i{\left|\frac{\left.a_{\mathcal{D}}\right|_{500}-\left.a_{\mathcal{D}} \right|_{n}}{\left.a_{\mathcal{D}}\right|_{500}} \right|_{t=t_i} }, \label{eq:ad_avg}\\
    \left<-q_{\mathcal{D}_n} \right> &=& \displaystyle \dfrac{1}{\sum{i}}\sum_i{\left|\left(\left.\frac{\ddot{a_{\mathcal{D}}}}{a_{\mathcal{D}} H_{\mathcal{D}}^2}\right|_{500}-\left.\frac{\ddot{a_{\mathcal{D}}}}{a_{\mathcal{D}} H_{\mathcal{D}}^2}\right|_{n}\right)\right|_{t=t_i}}.\label{eq:qd_avg}
\end{eqnarray}
The terms $\left<a_{\mathcal{D}_n} \right>$ and $\left<-q_{\mathcal{D}_n} \right>$ denote the time-averaged variation of $\left.a_{\mathcal{D}}\right|_n$ and $\left.\frac{\ddot{a_{\mathcal{D}}}}{a_{\mathcal{D}} H_{\mathcal{D}}^2}\right|_n$ respectively, from the limiting case of $n=500$. In this analysis, we split the entire time range (\emph{i.e.} $0 \leq t/t_0\leq 6$) into 1000 bins. In \autoref{eq:ad_avg} and \autoref{eq:qd_avg}, $i$ is the index number of such bins. The variation of $\left<a_{\mathcal{D}_n} \right>$ and $\left<-q_{\mathcal{D}_n} \right>$ with $n$ are plotted in \autoref{fig:aD_qD_n_n}(a) and \autoref{fig:aD_qD_n_n}(b) respectively. In these two figures, the plots are for different chosen sets of $\mu_o$ and $\mu_u$, while the other two parameters are kept at $\sigma_u=0.1$ and $\sigma_o=0.01$. It is clearly seen that for $n\geq 100$, the average fluctuation is less than $\sim 0.01$. For higher values of $\mu_u$, the fluctuations reduce further for even smaller values of $n$. In view of the above results, we chose $n=100$ for our remaining calculations.

\begin{figure*}
    \centering
    \begin{tabular}{cc}
	\includegraphics[width=0.45\textwidth]{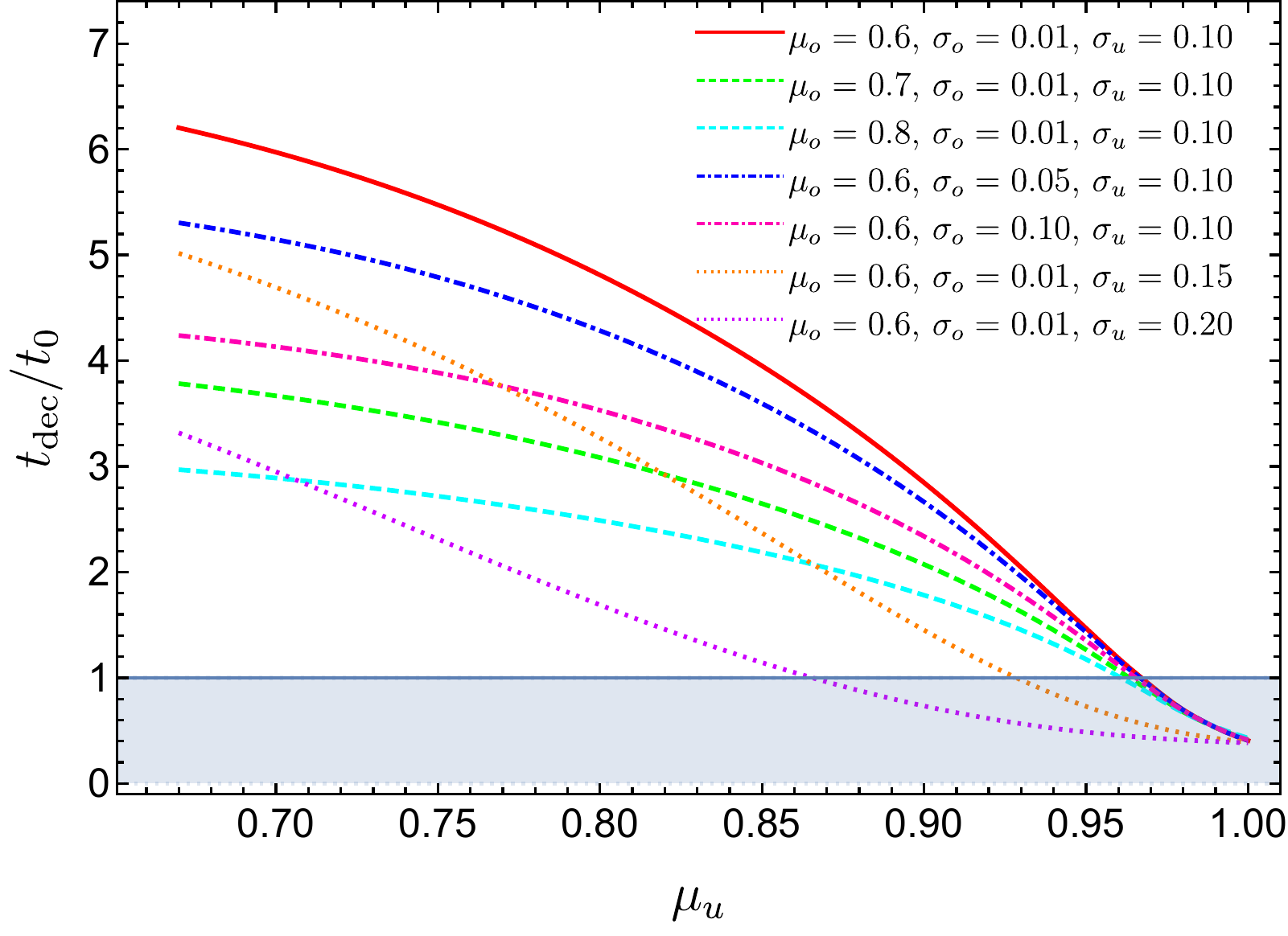}&
	\includegraphics[width=0.45\textwidth]{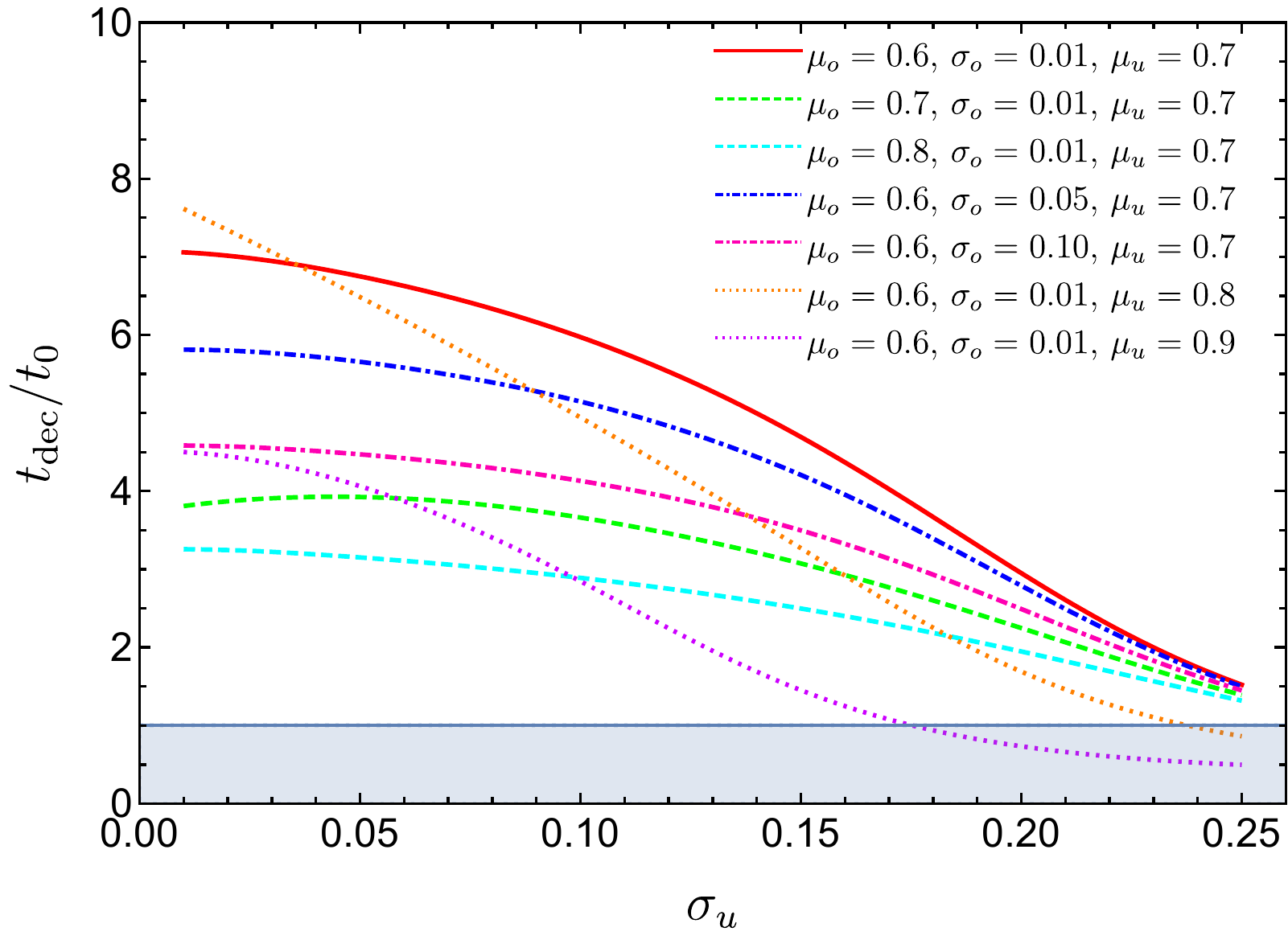}\\
	(a)&(b)\\
	\includegraphics[width=0.45\textwidth]{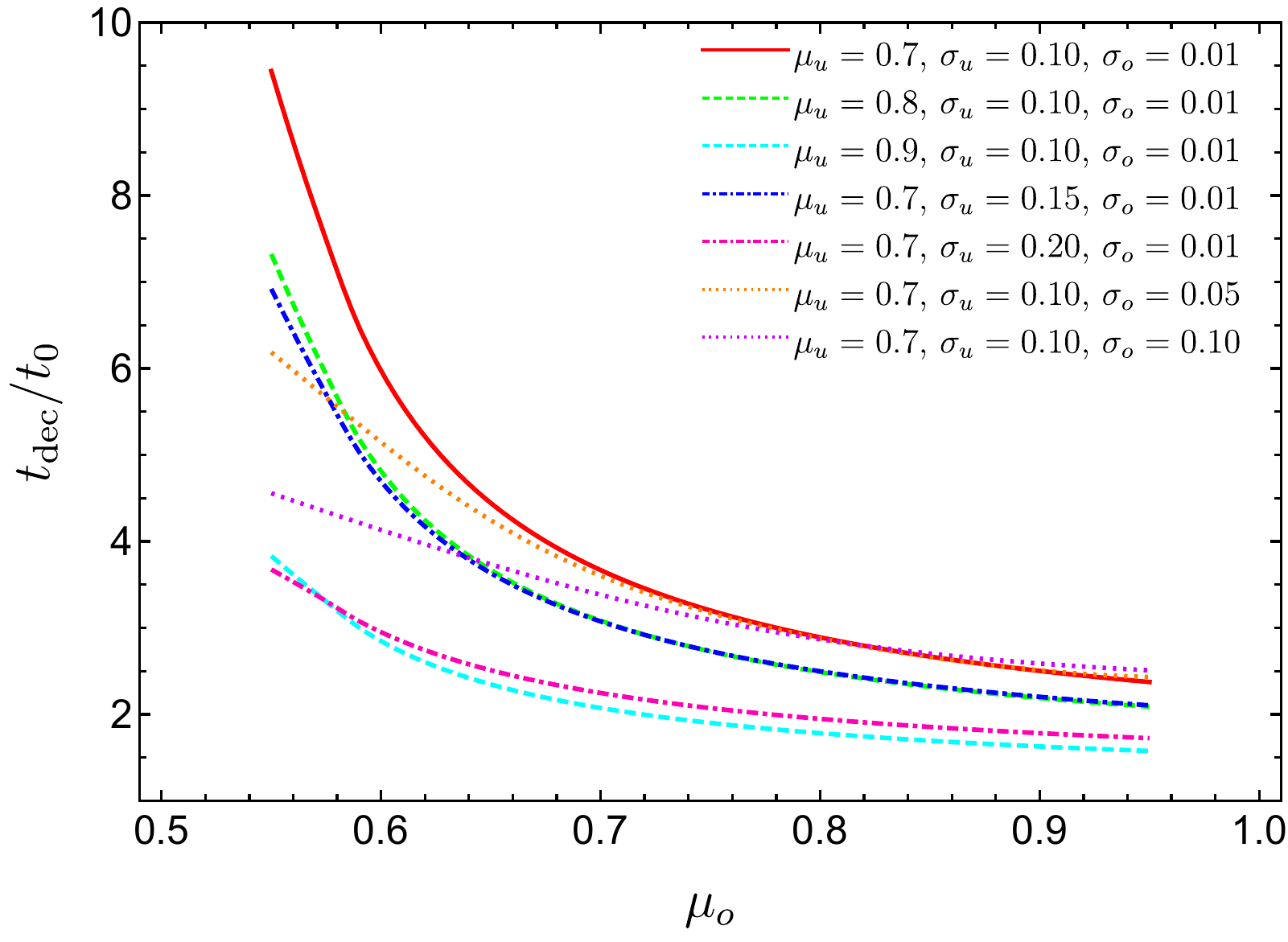}&
	\includegraphics[width=0.45\textwidth]{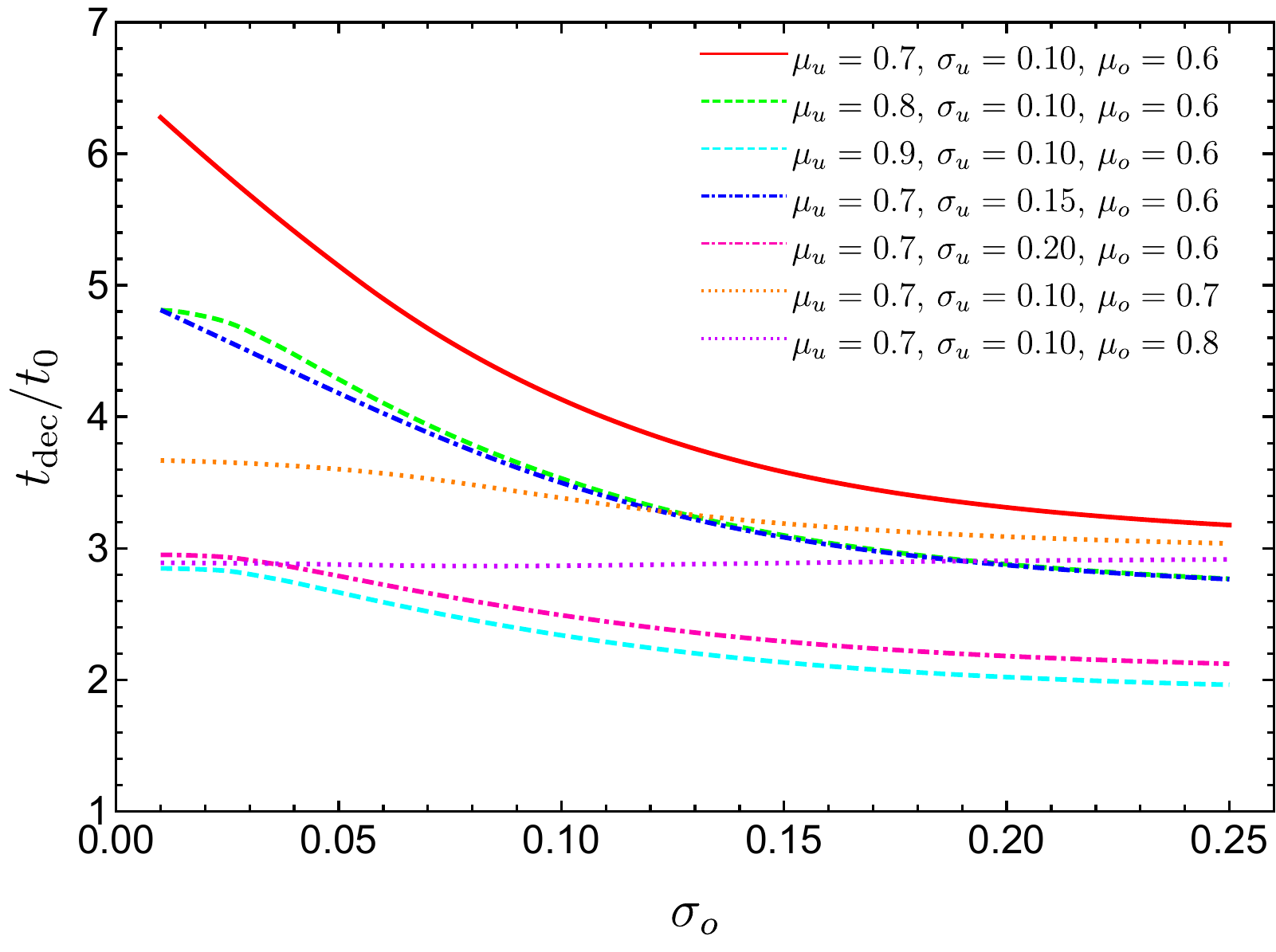}\\
	(c)&(d)\\
    \end{tabular}
    \caption{\label{fig:tdec} Variation of $t_{\rm dec}$ with (a) $\mu_u$ for different chosen values of $\mu_o$, $\sigma_o$ and $\sigma_u$, (b) $\sigma_u$ for different chosen values of $\mu_o$, $\sigma_o$ and $\mu_u$, (c) $\mu_o$ for different chosen values of $\mu_u$, $\sigma_u$ and $\sigma_o$, (d) $\sigma_o$ for different chosen values of $\mu_u$, $\sigma_u$ and $\mu_o$.}
\end{figure*}
	
In the present analysis, we essentially look into the late-time dynamics of the Universe. As in \autoref{fig:aD_qD}(b) and \autoref{fig:qD_n}, one can see that, although the Universe is expanding at the current epoch, the rate of acceleration $\left(\frac{\ddot{a_{\mathcal{D}}}}{a_{\mathcal{D}} H_{\mathcal{D}}^2}\right)$ will start falling in the future, and after a certain time ($t_{\rm dec}$), the quantity $\frac{\ddot{a_{\mathcal{D}}}}{a_{\mathcal{D}} H_{\mathcal{D}}^2}$ exhibits negative values, depicting the deceleration of the Universe. The value of $t_{\rm dec}$ varies  with all four model parameters \emph{i.e.} $\mu_u$, $\sigma_u$, $\mu_o$ and $\sigma_o$. In \autoref{fig:tdec}(a), the variation of $t_{\rm dec}$ with $\mu_u$ is graphically represented for various sets of the other three parameters (\emph{i.e.} $\sigma_u$, $\mu_o$ and $\sigma_o$). Similar variations with $\sigma_u$, $\mu_o$ and $\sigma_o$ are shown in \autoref{fig:tdec}(b), \autoref{fig:tdec}(c) and \autoref{fig:tdec}(d) respectively. From these figures it can be seen that, with increase of the model parameters, we get lower values of $t_{\rm dec}$.  An increase in $\mu_o$ or $\mu_u$, results in the increase of the most frequent value of $q_{o_i}$ or $\beta_i$ in the distribution of overdense or underdense subdomains,  respectively. Hence, higher values of $\mu_o$ result in overdense subdomains with higher values of $q_{o_i}$ becoming more prominent which leads to the acceleration parameter falling faster (refer to the discussion for \autoref{fig:aD_qD}) and hence, results in lower values of $t_{\rm dec}$. Similarly, higher values of $\mu_u$ result in underdense subdomains with higher values of $\beta_i$ becoming more prominent leading to lower values of $t_{\rm dec}$. Higher values of $\sigma_o$ for a given $\mu_o$ result in the distribution of subdomains becoming wider around the mean value of $\mu_o$. Therefore, there are now more subdomains with a larger value of $q_{o_i}$ in the distribution, resulting in lower values of $t_{\rm dec}$. Similarly, higher values of $\sigma_u$ for a given $\mu_u$ result in more subdomains with a larger value of $\beta_i$ in the distribution resulting in lower values of $t_{\rm dec}$.
In \autoref{fig:tdec}(a) and \autoref{fig:tdec}(b), one can see that for higher values of $\mu_u$ in the presence of higher $\sigma_u$, the values of $t_{\rm dec}$ is less than $t_0$, which contradicts with observational evidence. Hence such values are ruled out of the permissible range (shaded region).    
	
\begin{figure*}
    \centering
    \begin{tabular}{cc}
	\includegraphics[trim={0 0 85 0},clip, width=0.45\textwidth]{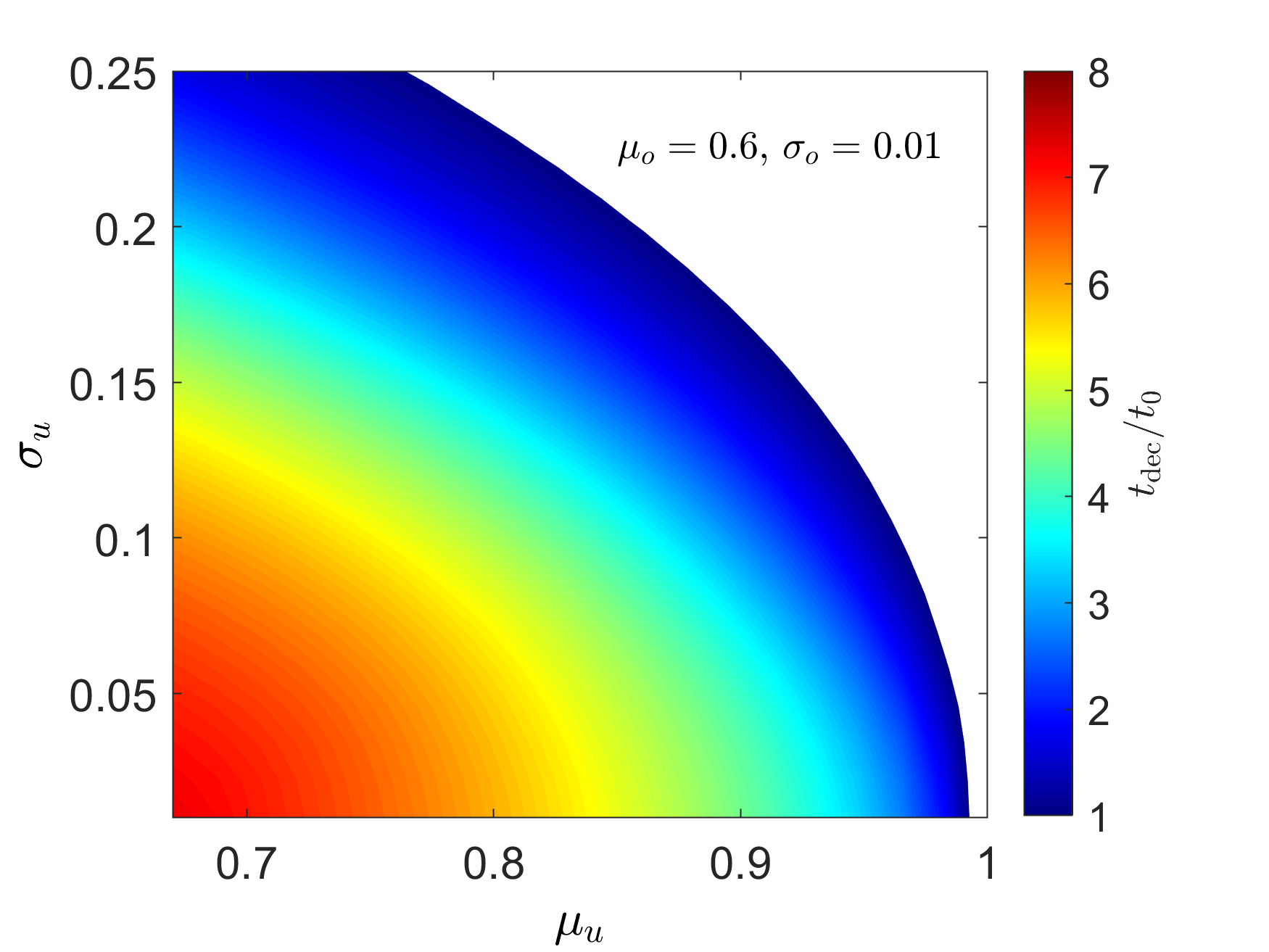}&
	\includegraphics[trim={0 0 85 0},clip, width=0.45\textwidth]{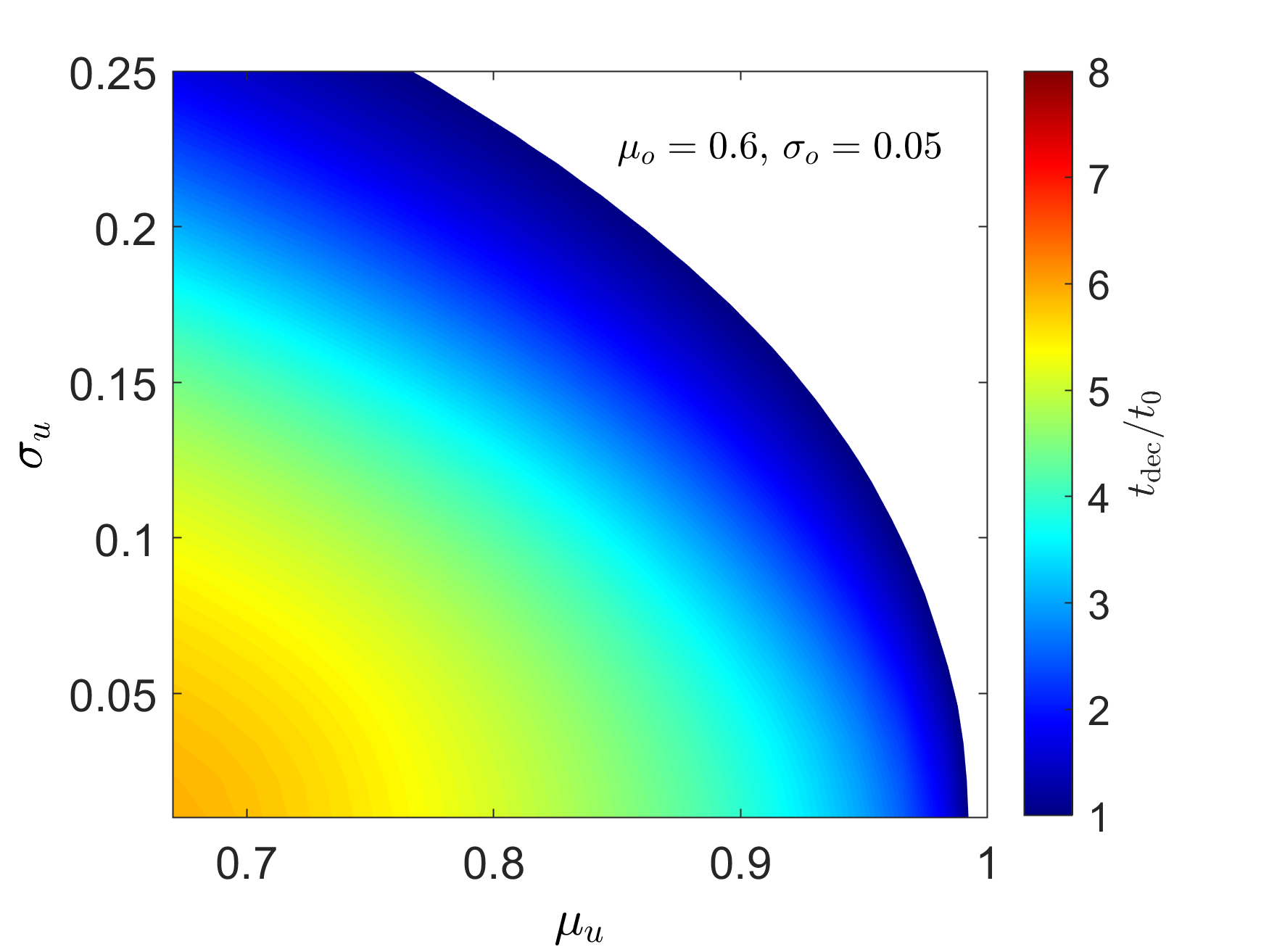}\\
	(a)&(b)\\
	\includegraphics[trim={0 0 85 0},clip, width=0.45\textwidth]{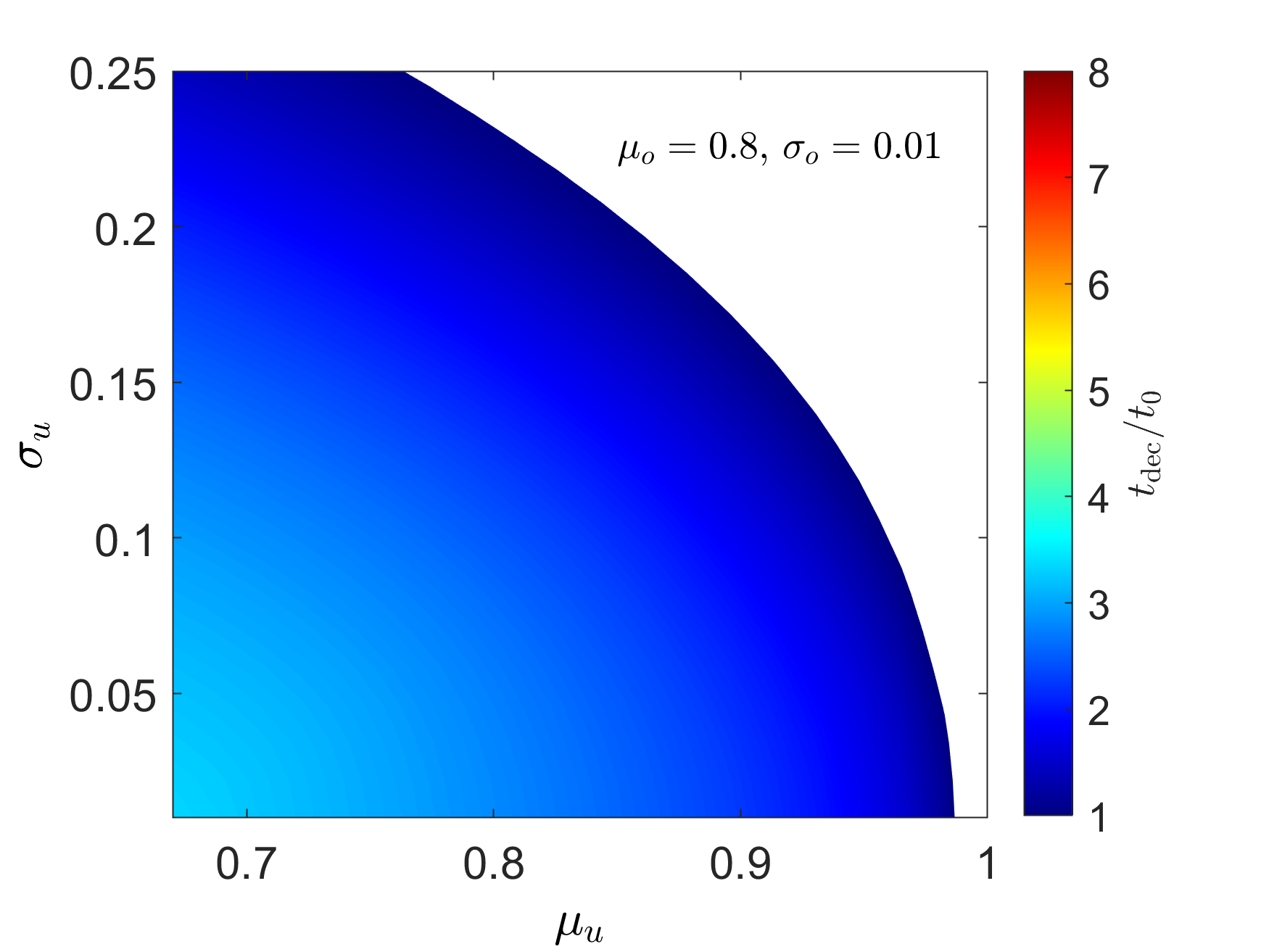}&
	\includegraphics[trim={0 0 85 0},clip, width=0.45\textwidth]{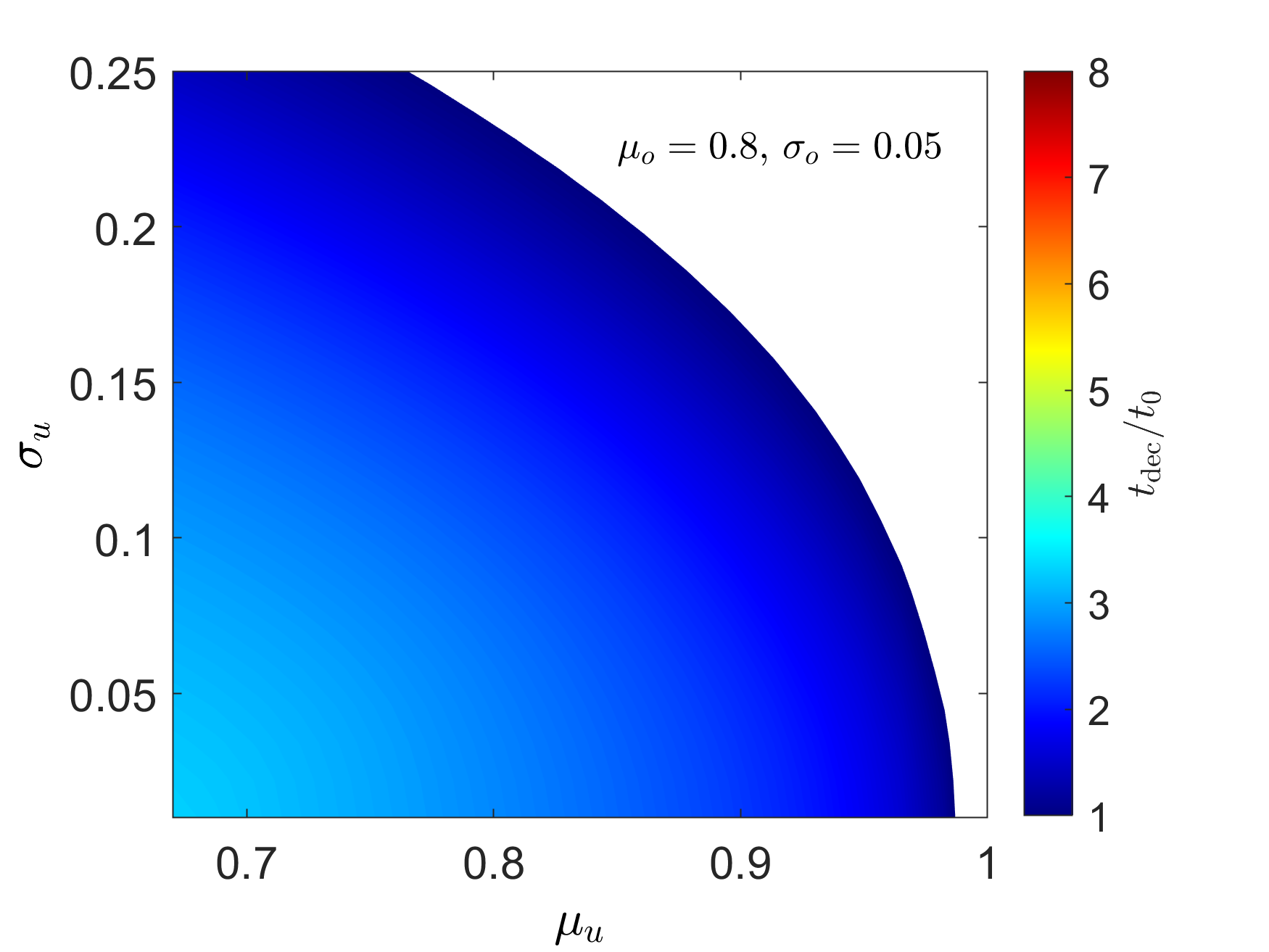}\\
	(c)&(d)\\
    \end{tabular}
    \begin{tabular}{c}
	\includegraphics[trim={0 0 0 230},clip, width=0.7\textwidth]{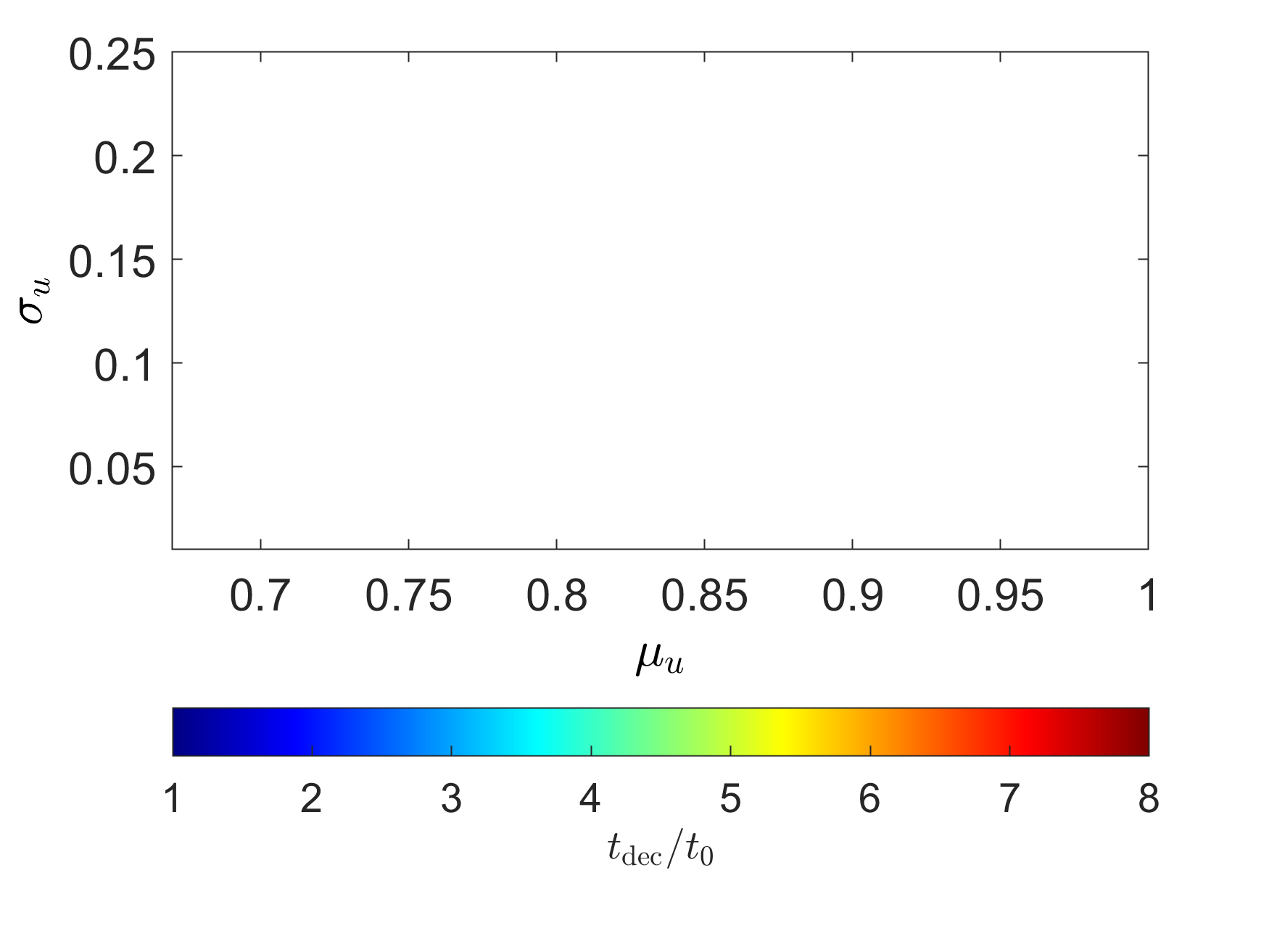}\\		
    \end{tabular}
    \caption{\label{fig:tdec_cont_o} Contour representation of $t_{\rm dec}$ in $\sigma_u - \mu_u$ plane at (a) $\mu_o=0.6$, $\sigma_o=0.01$ (b) $\mu_o=0.6$, $\sigma_o=0.05$, (c) $\mu_o=0.8$, $\sigma_o=0.01$, (d) $\mu_o=0.8$, $\sigma_o=0.05$.}
\end{figure*}

In \autoref{fig:tdec_cont_o}, the variation of $t_{\rm dec}$ in $\mu_u - \sigma_u$ plane is shown for different sets of $\mu_o$ and $\sigma_o$. The values of $t_{\rm dec}$ are described using different colours, as mentioned in the colour bar furnished at the bottom of \autoref{fig:tdec_cont_o}. Here, the white regions denote the area where $t_{\rm dec}\leq t_0$ is beyond the permissible range. From \autoref{fig:tdec_cont_o} one can see that, the value of $t_{\rm dec}$ falls sharply with respect to the point  $(\mu_u,\sigma_o) = (0.67,0.01)$ in the scaled axes ($\mu_u \in [0.67, 1.00]$, $\sigma_u \in [0.01, 0.25]$). The plot of \autoref{fig:tdec_cont_o}(a) represents the case of $\mu_o=0.6$, $\sigma_o=0.01$, where the maximum possible value of $t_{\rm dec}\approx 7.5 t_0$ is obtained at $\mu_u=0.67$, $\sigma_o = 0.01$. But it decreases remarkably at a higher value of $\sigma_o$ ($\sigma_o=0.05$) (comparing parts (a) and (b) of the figure). $t_{\rm dec}$ also decreases  on increasing the value of $\mu_o$ as can be seen by comparing parts (a) and (c) of the figure where $\mu_o$ has undergone an  increased from $0.6$ to $0.8$. This is in accordance with our previous analysis of \autoref{fig:tdec} and the related discussion showing that $t_{\rm dec}/t_0$ has lower values for higher values of $\mu_o$ and $\mu_u$. \autoref{fig:tdec_cont_o}(c) and \autoref{fig:tdec_cont_o}(d) are the same to the plots of \autoref{fig:tdec_cont_o}(a) and \autoref{fig:tdec_cont_o}(b) respectively, where the parameter $\mu_u$ is set at 0.8. Comparing those plots, it can be seen that, at higher values of $\mu_o$, the variation of $t_{\rm dec}$ is comparatively small and $t_{\rm dec}/t_0$ is very close to unity ($t_{\rm dec} \lessapprox 3 t_0 $ for $\mu_o=0.8$). It can also be seen from parts (c) and (d) of the figure that the variation of $t_{\rm dec}$ is also negligible on increasing the value of $\sigma_o$ for such higher values of $\mu_o$.

\begin{figure*}
    \centering
    \begin{tabular}{cc}
	\includegraphics[trim={0 0 80 0},clip, width=0.45\textwidth]{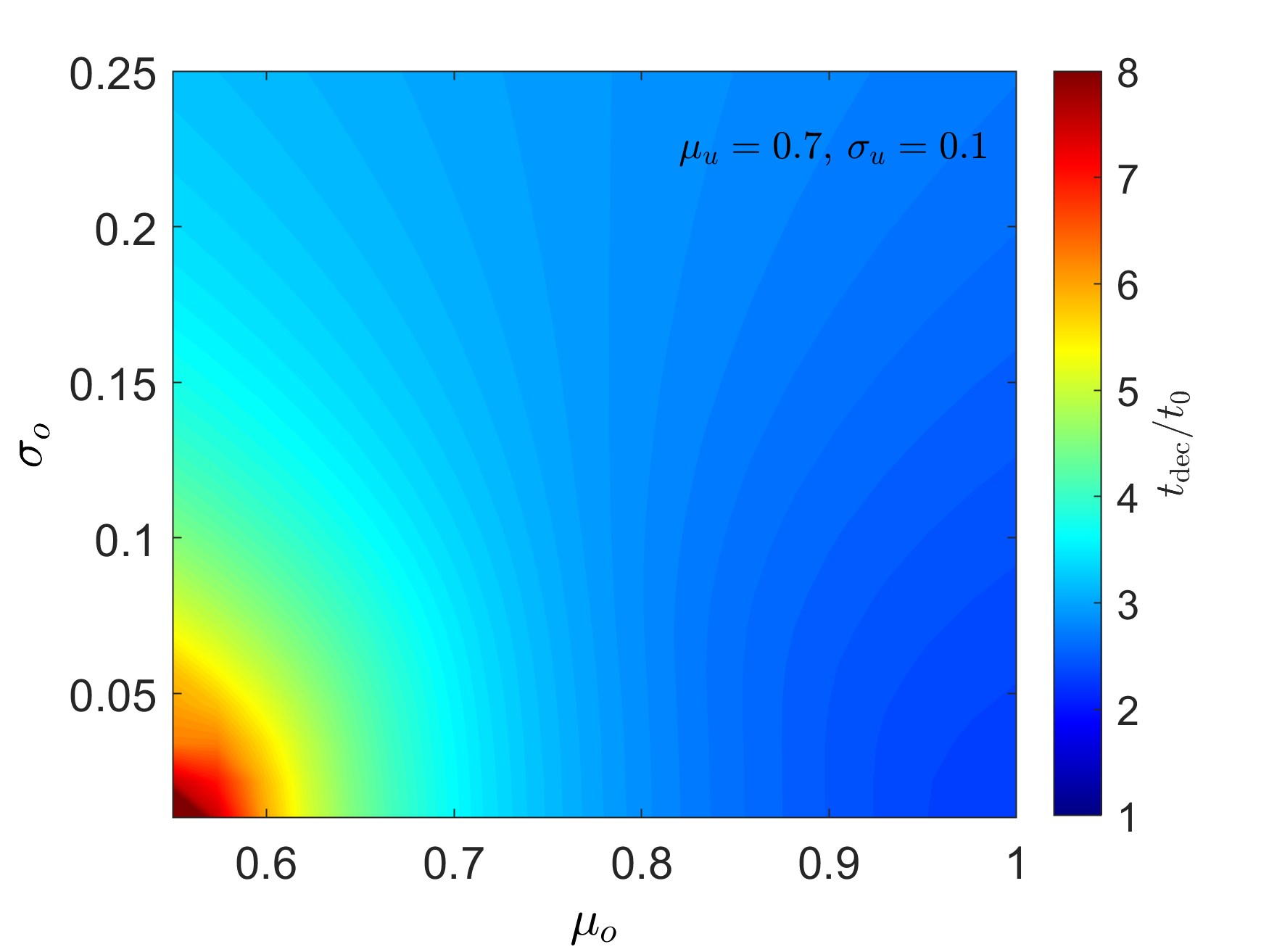}&
	\includegraphics[trim={0 0 80 0},clip, width=0.45\textwidth]{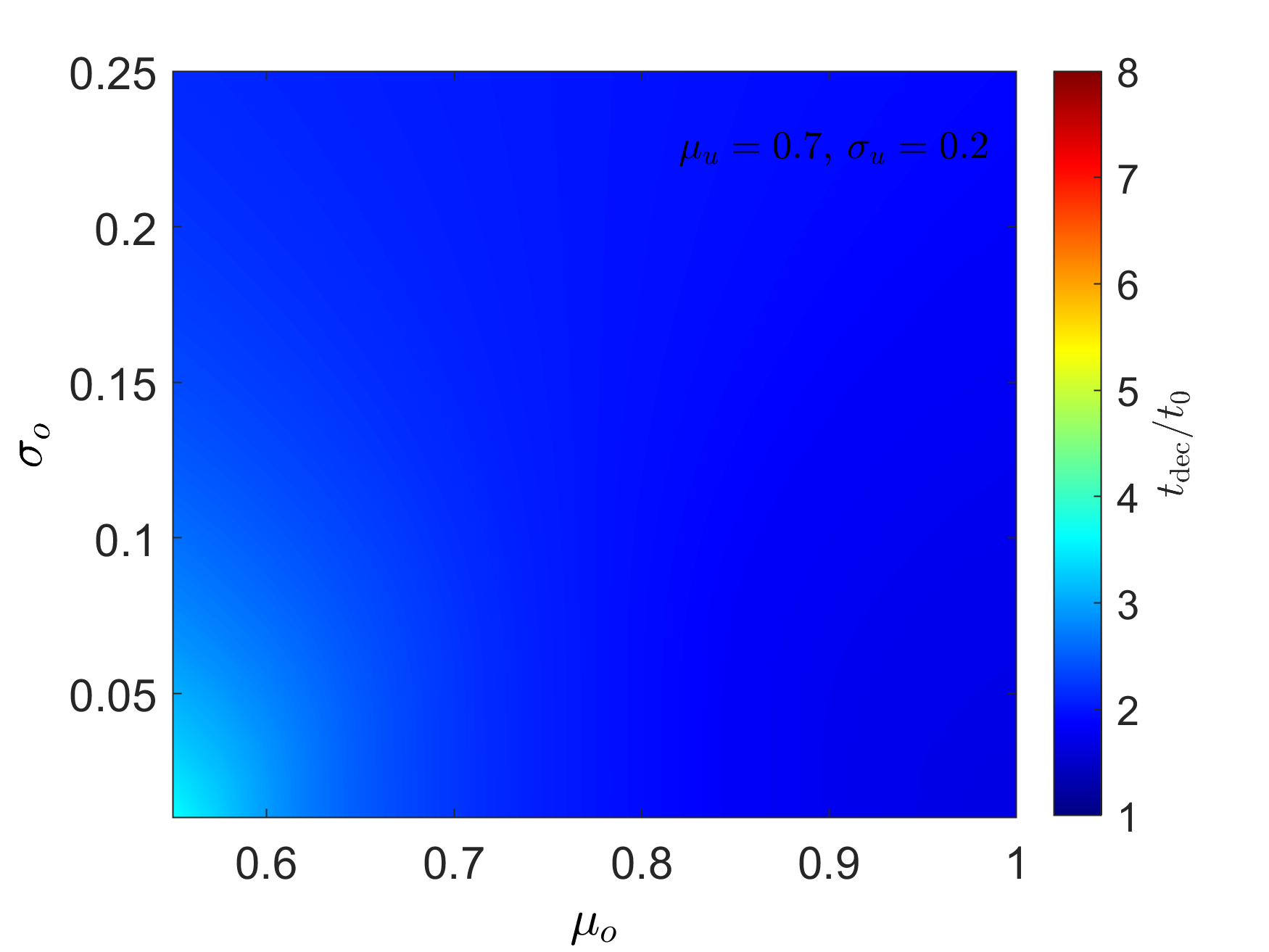}\\
	(a)&(b)\\
	\includegraphics[trim={0 0 80 0},clip, width=0.45\textwidth]{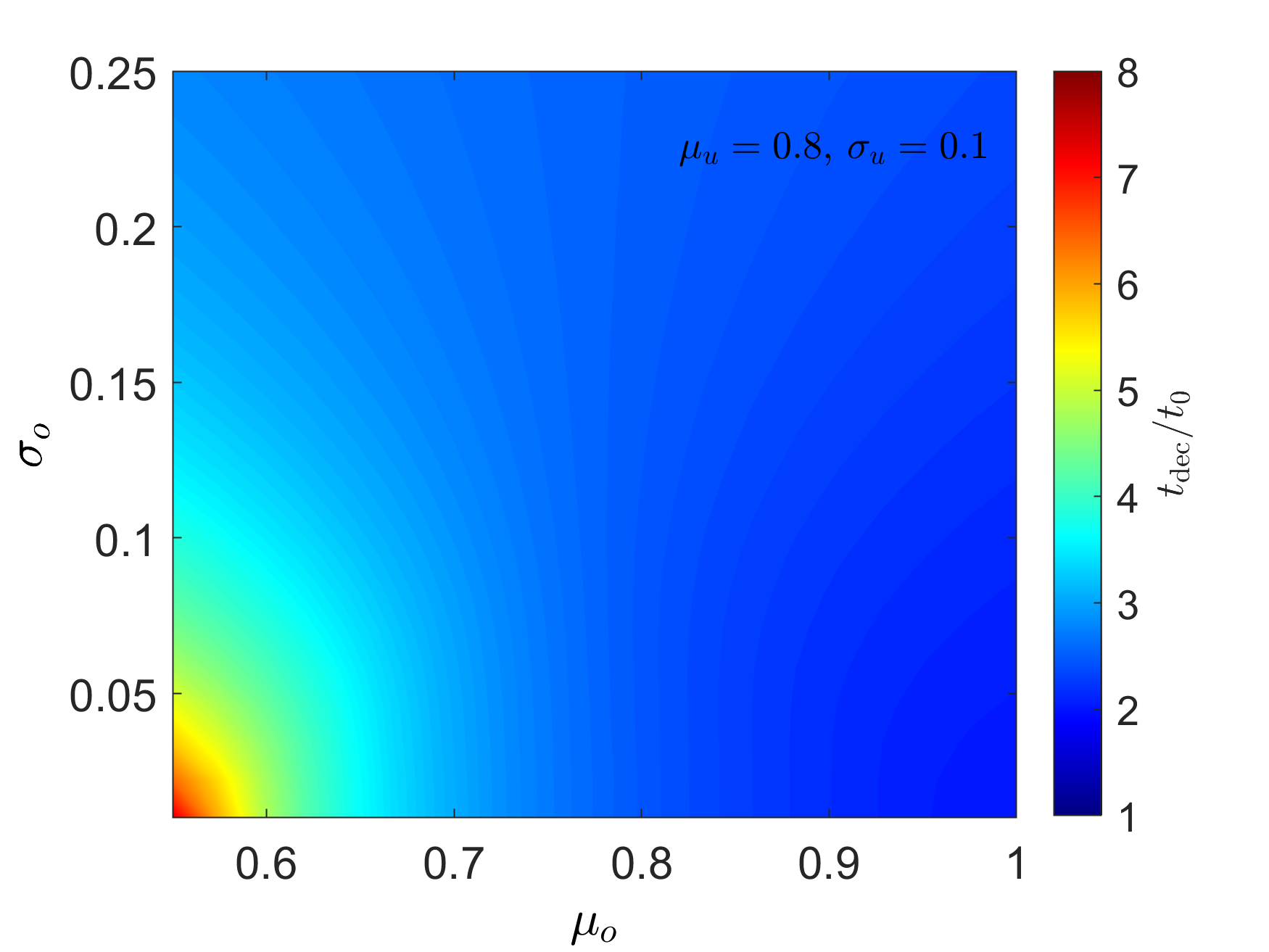}&
	\includegraphics[trim={0 0 80 0},clip, width=0.45\textwidth]{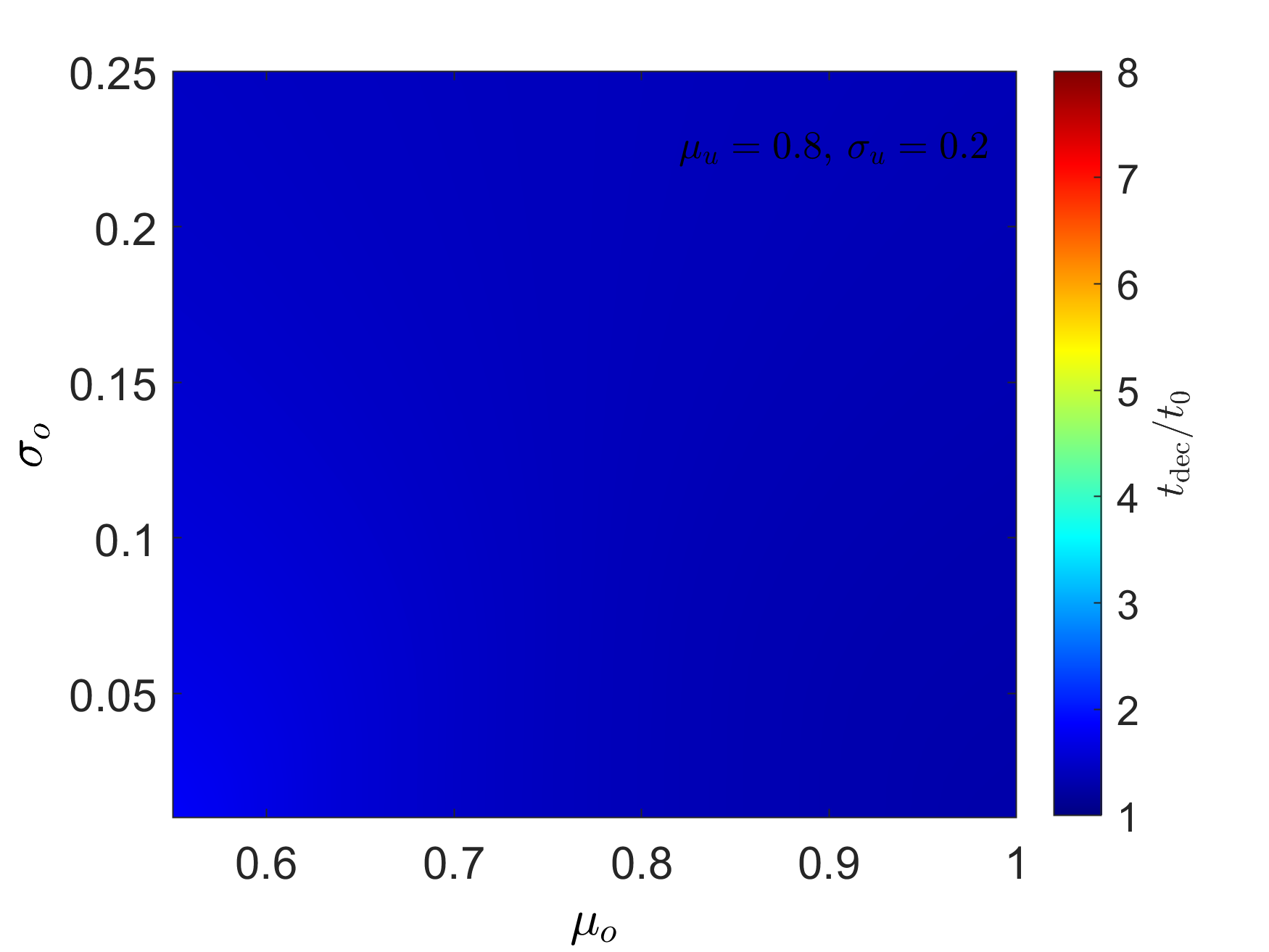}\\
	(c)&(d)\\
    \end{tabular}
    \begin{tabular}{c}
	\includegraphics[trim={0 0 0 230},clip, width=0.7\textwidth]{figures/contour_tdec_cbar.png}\\		
    \end{tabular}
    \caption{\label{fig:tdec_cont_u} Contour representation of $t_{\rm dec}$ in $\sigma_o - \mu_o$ plane at (a) $\mu_u=0.7$, $\sigma_u=0.1$ (b) $\mu_u=0.7$, $\sigma_u=0.2$, (c) $\mu_u=0.8$, $\sigma_u=0.1$, (d) $\mu_u=0.8$, $\sigma_u=0.2$.}
\end{figure*}

\par A similar analysis is performed in the $\sigma_o - \mu_o$ plane. In \autoref{fig:tdec_cont_u}(a), the contour representation shown the variation of $t_{\rm dec}$ for the case of $\mu_u=0.7$ and $\sigma_u=0.1$. \autoref{fig:tdec_cont_u}(b) is plotted for $\mu_u=0.7$ and $\sigma_u=0.2$. \autoref{fig:tdec_cont_u}(c) and \autoref{fig:tdec_cont_u}(d) are the same to the plots \autoref{fig:tdec_cont_u}(a) and \autoref{fig:tdec_cont_u}(b) respectively, but for $\mu_u=0.8$. Larger values of $t_{\rm dec}/t_0$ are obtained in part (a) of the figure for lower values of $\mu_o$ and $\sigma_o$. As the values of $\mu_o$ and $\sigma_o$ are increased, the value of $t_{\rm dec}/t_0$ decreases. From parts (a) and (c), one can notice that for higher values of $\mu_u$, the value of $t_{\rm dec}/t_0$ decreases if the values of ($\mu_o$, $\sigma_o$) are kept fixed.  The variation of $t_{\rm dec}$ 
becomes insignificant for higher values of the parameters. Parts (b) and (d) show that at higher values of $\sigma_u$, the variation in the $\sigma_o - \mu_o$ plane almost vanishes. These results are in accordance with the analysis and discussion relating to \autoref{fig:tdec}.

\section{Observational constraints}\label{sec:obs}

\begin{figure*}
    \centering
    \includegraphics[width=\textwidth]{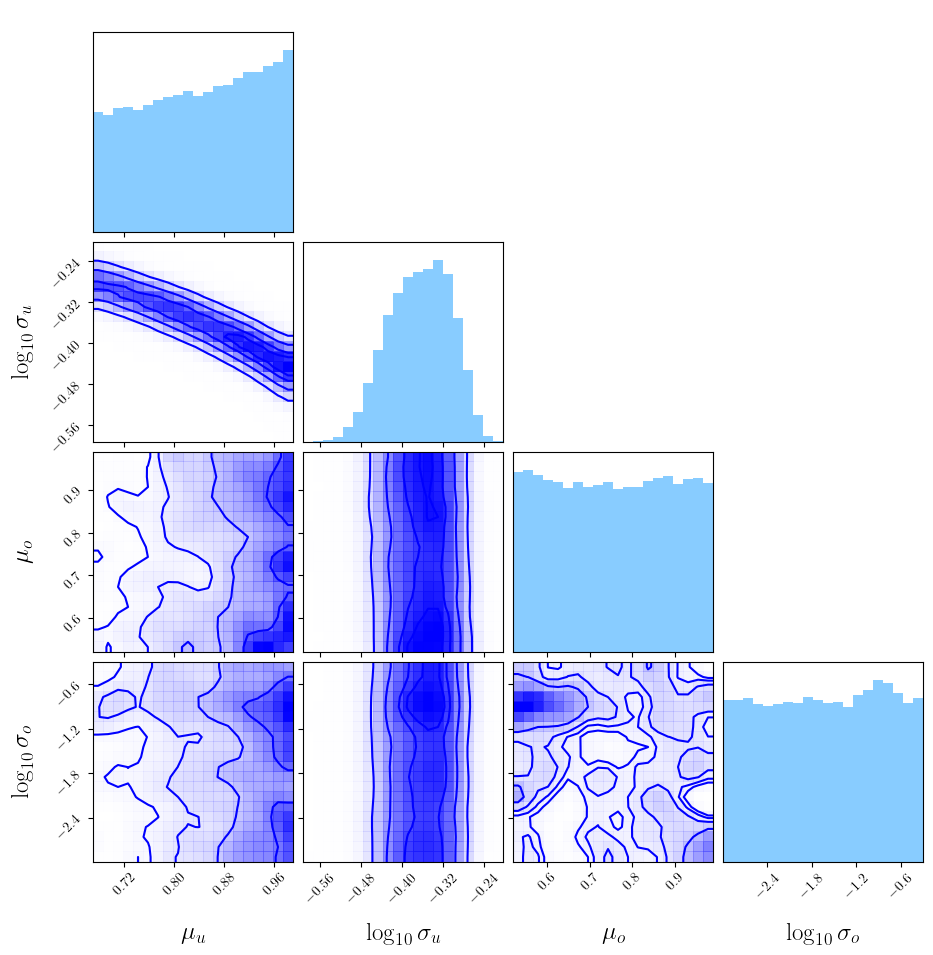}
    \caption{\label{fig:union} Corner plot showing the MCMC result for our model carried out using the observational results of Union 2.1 supernova Ia data \cite{union}. The histograms on the diagonal show the marginalized posterior densities for each parameter. }
\end{figure*}

In \autoref{fig:aD_qD} - \autoref{fig:tdec_cont_u}, we have utilized our model to examine the future evolution of an inhomogeneous multi-domain-ed spacetime, and  analyzed the variation of various cosmological quantities with respect to our model parameters. Now, we examine our model with respect to observational data and determine the optimum values of our model parameters. We carry out a Bayesian analysis in order to compare our model with Union 2.1 supernova Ia data \cite{union}. We  use the distance modulus versus redshift data from Union 2.1. In order to compare our model with the observational data, we need a scheme to relate the theoretically calculated quantities from our model with observational quantities. For this purpose, we  employ the covariant scheme \cite{rasanen1, rasanen2}. The covariant scheme gives us a relation between effective redshift and the angular diameter distance, given by

\begin{align}
    1+z = \frac{1}{a_D}\label{eq:covariant_sch_1}\\
    H_D\frac{d}{dz}\left((1+z)^2H_D\frac{dD_A}{dz}\right) = -4\pi G\langle\rho\rangle_D D_A.\label{eq:covariant_sch_2}
\end{align}
The angular diameter distance $D_A$ can, in turn, be transformed into the distance modulus as a function of redshift for our model using standard cosmological distance relations. 

\par In this analysis, the resulting posterior distributions of different parameters are obtained by Markov Chain Monte Carlo (MCMC) iteration method (\autoref{fig:union}) by using the \texttt{MCMCSTAT} package \cite{mcmc1,mcmc2}. We use total $10^5$ number of events with the adaptation interval of $300$, within the parameter range: $\mu_u \in [0.67, 1.00]$, $\log_{10}\sigma_u \in [-0.60, -0.15]$, $\mu_o \in [0.51, 1.00]$ and $\log_{10}\sigma_o \in [-3.00, -0.30]$. Initially, the same analysis was carried out with a wider range of $\log_{10}\sigma_u$ and $\log_{10}\sigma_o$, but later those ranges were redefined in order to skip the extremely lower posterior regions. 

In \autoref{fig:union}, the parameters $\sigma_u$ and $\sigma_o$ are shown in log-scale while the other two parameters $\mu_u$ and $\mu_o$ are shown in linear-scale. The topmost plots of the first, second, third and fourth column of \autoref{fig:union} represent the posterior distribution for the parameters $\mu_u$, $\log_{10}\sigma_u$, $\mu_o$ and $\log_{10}\sigma_o$ respectively, while the other plots of \autoref{fig:union} show the contour representation of the posterior distribution in different sets of two-parameter space. In these contour plots, the regions with darker colours denote higher posterior regions, and the lines indicate the boundaries of $1\sigma$, $2\sigma$ and $3\sigma$ regions, respectively. The posterior plots essentially describe the epistemic uncertainties of the corresponding model parameters. The diagonal panels show the 1-D histogram of the posterior distribution for each model parameter obtained by marginalizing the other parameters. The off-diagonal panels show 2-D projections of the posterior probability distributions for each pair of parameters and correlations between the parameters, with contours.

From this analysis the obtained set of optimum points are $\mu_u = 0.85^{+0.10}_{-0.12}$, $\log_{10}\sigma_u = -0.36^{+0.06}_{-0.07}$, $\mu_o = 0.75^{+0.16}_{-0.16}$ and $\log_{10}\sigma_o = -1.63^{+0.90}_{-0.94}$ respectively. However, from the posterior plot for $\log_{10}\sigma_u$ (in \autoref{fig:union}), one can notice that the highest probable value (or best-fit value) of $\log_{10}\sigma_u$ is slightly higher ({\emph i.e.} $\log_{10}\sigma_u=-0.33$) than the corresponding optimum point. On the other hand, the best-fit point for $\mu_u$ lies on its highest posterior value (\emph{i.e.} $\mu_u=1.00$). From the histograms of \autoref{fig:union}, it is evident that variation of the parameters of the overdense regions ($\mu_o$ and $\sigma_o$) do not play a very significant role as histograms depicting the marginalized posterior distributions for these two parameters do not show any specific significant trend. Similarly, in the off-diagonal panels of these two parameters, there is no specific trend, while the off-diagonal panels of other pairs show some specific trend. For example, the $\mu_u-log_{10}\sigma_o$ and $\mu_u - \mu_o$ panels show a specific trend towards higher values of $\mu_u$ (darker regions are towards higher values of $\mu_u$ and this trend is also indicated by the histogram for $\mu_u$). These indicate that the modification of the volume fractions distribution of the overdense regions by modifying the mean and standard deviation of the distribution (\autoref{eq:gauss_o}) doesn't have much effect on the future evolution of the Universe, as the fraction of overdense region at the present era is taken to be $\sim 9 \%$ \cite{Weigand_et_al}.


\section{Conclusions}\label{sec:conclusions}

In this work, we have considered a multi-domain model of spacetime having inhomogeneous matter distribution. The multiple subregions in our model are broadly categorized into two types - overdense and underdense, with all such subregions having distinct evolution parameters.  In the context of the Buchert formalism, we have computed the averaged backreaction of the matter inhomogeneities
on the late time global evolution of the Universe.  

Our results clearly indicate that the global acceleration falls with time beyond the present epoch for a significant range of values of our model parameters. Such a feature predicted earlier for dust Universe models \cite{Constraints_Räsänen_2006}, and also observed in the context of simplified
two-scale models \cite{bose, Bose2013, Ali_2017}  is further corroborated here
through the analysis of a more realistic model. We have shown here that
after a particular time $(t_{dec})$, the value of the global acceleration parameter can become negative, signifying the transition of the presently
accelerating Universe to a phase of future deceleration. 

Through our analysis, we have optimized our model for the maximum
number of subregions to be considered for a reliable result
for future global evolution. The dependence of the deceleration time $(t_{dec})$ on the various model parameters has been analyzed systematically.  The variation of $(t_{dec})$ shows that the model parameters associated with the underdense subregions have more impact on the transition time for
future deceleration. We have further correlated our model with observation data. We have obtained the marginalised posterior densities for each model parameter through Markov Chain Monte Carlo (MC-MC) simulations using the Union 2.1 supernova Ia data. 

We conclude by noting that though the present era Universe is accelerating \cite{Perlmutter1998, Riess_1998, Hicken_2009, Marina_Seikel_2009}, such
behaviour could indeed be transitory. Observations have shown that the present Universe has an inhomogeneous matter distribution at considerably large scales \cite{Labini_2009, wiegand_scale}. It seems inevitable for an impact of backreaction of matter inhomogeneities on the global metric to plausibly avoid the future big  chill in the $\Lambda$CDM model or a possible future big rip problem in 
the presence of phantom dark energy. Our present results motivate further investigations in the context of various backreaction schemes and upcoming probes with more accurate observations to critically examine our conjecture of the future deceleration of the Universe.

\section{Acknowledgements}
The authors would like to thank Amna Ali for the discussions. SSP would like to thank the Council of Scientific and Industrial Research (CSIR), Govt. of India, for funding through the CSIR-SRF-NET fellowship.

\bibliography{references}
\end{document}